\documentclass[sigconf, nonacm, pdfa]{acmart}

\usepackage{xspace}

\newcommand\vldbdoi{XX.XX/XXX.XX}
\newcommand\vldbpages{XXX-XXX}
\newcommand\vldbvolume{14}
\newcommand\vldbissue{1}
\newcommand\vldbyear{2020}
\newcommand\vldbauthors{\authors}
\newcommand\vldbtitle{\shorttitle} 
\newcommand\vldbavailabilityurl{https://github.com/AgenticDataBench/AgenticDataBench}
\newcommand\vldbpagestyle{plain}

\usepackage{graphicx}
\usepackage{balance}  
\usepackage{epsfig}
\usepackage{epstopdf}
\usepackage{latexsym}
\usepackage{enumitem}
\PassOptionsToPackage{noend}{algpseudocode}
\usepackage{algpseudocode}
\usepackage{pifont}
\usepackage{epsfig,endnotes}
\usepackage{amssymb}
\usepackage{amsmath}
\usepackage{amsfonts}
\usepackage{stmaryrd}
\usepackage{url}
\usepackage{multirow}
\usepackage{array}
\usepackage[normalem]{ulem}
\usepackage{color}
\usepackage{cleveref}
\usepackage{xspace}
\usepackage{mathtools}
\usepackage{soul}
\usepackage{listings}
\usepackage{xcolor}
\usepackage{tikz}
\newsavebox{\blackball}
\newsavebox{\greenball}

\usepackage[utf8]{inputenc}

\usepackage{bbding}
\usepackage{booktabs}
\usepackage{threeparttable}
\usepackage{tabulary}
\usepackage{CJKutf8}
\usepackage[most,breakable]{tcolorbox}
\usepackage{etoolbox}
\usepackage{fancyhdr}
\usepackage{lipsum}
\usepackage[fencedCode]{markdown}
\usepackage{hyperref}
\usepackage{makecell}

\usepackage{subcaption}
\usepackage{pifont}

\usepackage{listings,upquote,soul,framed}
\colorlet{shadecolor}{gray!20}

\usepackage[lined,boxed,vlined,ruled,linesnumbered]{algorithm2e}

\newcommand{\hi}[1]{\vspace{.25em} \noindent {\bf #1}\xspace}

\newcommand{\blue}[1]{\textcolor{blue}{#1}}

\newcommand{\oursys}{\ensuremath{\textit{AgenticDataBench}}\xspace}
\newcommand{\daagent}{\emph{DA-Agent}\xspace}
\newcommand{\smolagents}{\emph{Smolagents}\xspace}
\newcommand{\claudecode}{\emph{Claude Code}\xspace}
\newcommand{\codex}{\emph{CodeX}\xspace}
\newcommand{\qwen}{\emph{Qwen3.5}\xspace}
\newcommand{\kimi}{\emph{Kimi-K2.5}\xspace}
\newcommand{\claude}{\emph{Claude 4.6}\xspace}

\begin{document}

\title{AgenticDataBench: A Comprehensive Benchmark for Data Agents}


\author{Zhaoyan Sun$^{1,2}$, Shan Zhong$^1$, Daizhou Wen$^2$, Jiaxing Han$^2$, Guoliang Li$^1$, Ying Yan$^2$, Peng Zhang$^2$, Yu Su$^2$, Xiang Qi$^2$, Baolin Sun$^2$, Chengyuan Yang$^2$, Tao Fang$^2$, Huaiyu Ruan$^2$}

\affiliation{$^1$ Tsinghua University, $^2$ Ant Digital Technologies, Ant Group}

\email{szy22@mails.tsinghua.edu.cn,liguoliang@tsinghua.edu.cn,fuying.yy@antgroup.com}

\pagestyle{plain}

\pagenumbering{arabic}


\begin{abstract}
Data science aims to derive actionable insights from heterogeneous raw data, unlocking the value of the massive amounts of data generated in modern society. Automating this process is essential to reducing labor-intensive efforts for data scientists and enabling scalable data-driven applications. Recently, large language model (LLM)-based data agents have emerged as a promising solution to automate data science workflows. However, the field lacks comprehensive benchmarks to rigorously evaluate these agents across diverse scenarios with fine-grained granularity. 
To address this gap, we propose \oursys, a comprehensive benchmark featuring {\it realistic tasks spanning diverse domains with fine-grained ground-truth labels}. This enables evaluations to capture the diversity and complexity of data science workflows and the detailed performance of agents. First, to cover diverse domains, we collect real datasets and tasks from 15 vertical domains, including 5 real-world B2B use cases from a leading fintech company. Second, to remove redundancy in real-world tasks and generate high-quality tasks for domains lacking real data, we introduce data science skills, recurring data-centric operational patterns (e.g., ``Handling Missing Data''), and quantify benchmark coverage by the number of skills included. Representative skills are extracted from large-scale task solutions on Stack Overflow using skill-aligned hierarchical clustering. 
Third, for real-world business tasks, we select task-solution pairs that maximize diversity in skill composition, ensuring broad coverage of practical scenarios. Fourth, to generate realistic tasks for devise domains without real tasks, we propose a systematic LLM-based task generation approach to create workflows and tasks based on these skills. Finally, we evaluate state-of-the-art data agents using our annotated benchmark and open-sourced testbed, providing detailed skill-level insights.
\end{abstract}

\vspace{-2em}

\maketitle

\vspace{-1em}
\section{Introduction}
\label{sec:intro}
Data science aims to extract actionable insights from heterogeneous raw data, which plays a central role in realizing the value of massive data generated in modern IT and business~\cite{gartner2025dsml}.
Traditionally, data scientists expend substantial effort on understanding and processing poorly organized data, incorporating implicit domain knowledge, and iteratively implementing complex codes.

Recent advancements of large language models (LLMs) have demonstrated superiority in data science–related tasks such as planing~\cite{wang2025unify,shankar2025docetl,hu2026opensql,shankar2026task}, reasoning~\cite{yang2025qwen3,guo2025deepseek,zhang2026reward}, database operations~\cite{zhou2024d,zhou2024db,sun2025r,sun2025d,zhou2026cracksql,xu2026prepbench,xu2026bridging,lyu2026genia,zhou2026can,zhou2025cracking,zhou2025dbaiops}, and code generation~\cite{zeng2026glm,DBLP:journals/corr/abs-2604-06231}, leading to the emergence of data agents that automate end-to-end insight extraction from raw data with minimal human intervention~\cite{sun2025agenticdata,sun2025data,tang2026workspacebench10benchmarkingai,zhou2026readyagentnativememorysystem,qu2026st,lan2026agenticscholar,luo2026data,wang2026data,DBLP:conf/cidr/LiuPSZ0ACSYSZCC26}.

\begin{figure}[!t]
  \centering
  \includegraphics[width=\linewidth]{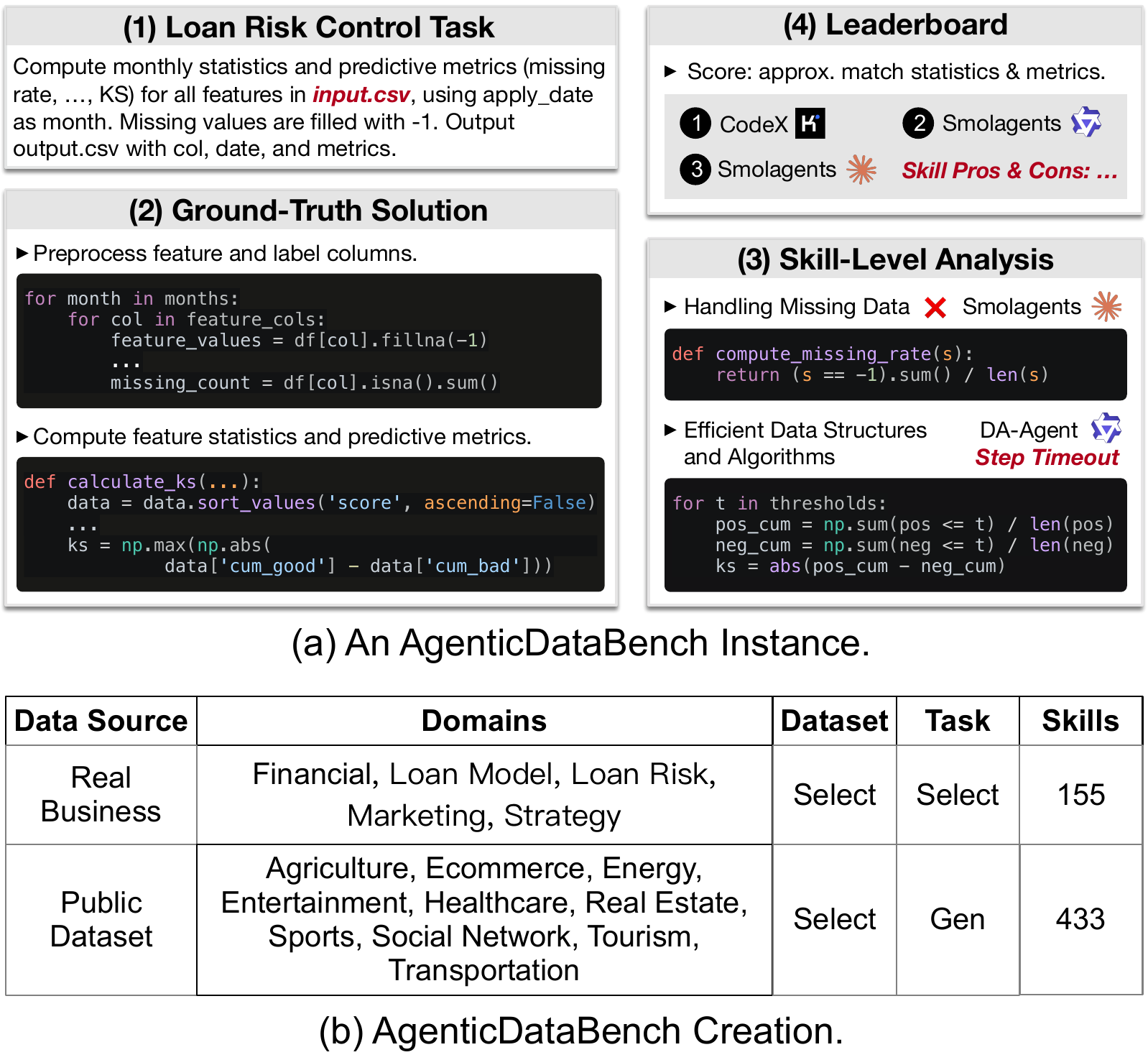}
  \vspace{-2.5em}
  \caption{Agentic Data Science Benchmark Example.}
  \label{fig:example}
  \vspace{-2em}
\end{figure}

Here we present a simplified workflow of data agents (see \autoref{fig:example}a).
$(i)$ \textit{Planning.} Given a complex data science task (e.g., predicting loan delinquency from monthly user statistics with AUC evaluation), the agent interprets user instructions and grounds them in relevant data sources. The challenge lies in instruction ambiguity (e.g., whether missing values should be filled with -1 instead of being pre-filled), heterogeneous data schemas, and large-scale datasets (e.g., ``input.csv'') that necessitate iterative exploration.
$(ii)$ \textit{Iterative Execution.} The agent iteratively plans actions, generates executable code, and interacts with execution environments (e.g., Python, databases). It progressively constructs an executable reasoning chain from intermediate results, such as adjusting feature processing or selecting more efficient algorithms under time constraints.
$(iii)$ \textit{Termination.} The process terminates upon either successful completion or reaching predefined step or time limits.

\hi{Motivation.} While numerous data agents have been proposed, a comprehensive evaluation framework for their systematic comparison is still lacking. Drawing from real-world data science practices, we identify that an effective benchmark should feature {\it realistic tasks spanning diverse domains, accompanied by fine-grained ground-truth labels}, enabling the evaluation to capture both the diversity and complexity of data science workflows as well as the detailed performance of agents. However, as shown in \autoref{tab:compare}, existing benchmarks fall short of meeting these criteria. They often rely on a limited set of manually selected task types, overlook the complexities of real-world business applications, and provide only coarse-grained task categories and aggregate scores, which obscure step-level behaviors.

\begin{table}[!t]
\caption{\oursys vs Existing Data Agent Benchmarks (-- indicates no solution code).}
\vspace{-1em}
\label{tab:compare}
\resizebox{\linewidth}{!}{
\begin{tabular}{|c|c|c|c|c|c|c|c|c|c|}
\hline
\textbf{Benchmark}           &  \textbf{\begin{tabular}[c]{@{}c@{}}\# Skills \\ Covered\end{tabular}} & \textbf{\begin{tabular}[c]{@{}c@{}}\# Tags \\ of Task\end{tabular}} & \textbf{\begin{tabular}[c]{@{}c@{}}\# Lines \\ of Code\end{tabular}} & \textbf{Data Source} & \textbf{Data Modality} & \textbf{\begin{tabular}[c]{@{}c@{}}Data (MB) \\ Per Task \end{tabular}} 
\\ \hline
DSBench~\cite{DBLP:conf/iclr/JingHWYYM0DY25}              & --                 & 2  & --                    & \begin{tabular}[c]{@{}c@{}}Public \\ Competition\end{tabular}      & \begin{tabular}[c]{@{}c@{}}(Semi-)Structured,  \\ Text\end{tabular}                 & 11.1                  
\\  \hline
BLADE~\cite{gu2024blade}              &   51                  & 2 & 16.3                      & Public Study                         & Structured                  & 2.4                  
\\  \hline
DA-Code~\cite{huang2024code}                          & --                   & 10 & 85~\cite{huang2024code}                  & Public Dataset              &  \begin{tabular}[c]{@{}c@{}}\begin{tabular}[c]{@{}c@{}} (Semi-)Structured,  \\ Markup, Text, Binary,  \\Database\end{tabular}\end{tabular}                & 23.0                 
\\  \hline
DataSciBench~\cite{zhang2025datascibench}        & 281                 & 6 & 33.4                      & Public Dataset                          & \begin{tabular}[c]{@{}c@{}} (Semi-)Structured,  \\ Text, Binary\end{tabular}                  & 0.8                  
\\  \hline
ScienceAgentBench~\cite{chen2025scienceagentbench}       & 220 & 11                  & 40.1                      & Public Study                          & \begin{tabular}[c]{@{}c@{}} (Semi-)Structured,  \\ Markup, Text, Binary\end{tabular}                   & 54.3                   
\\  \hline
KramaBench~\cite{lai2025kramabench}          & 194               & 9 & 34.4                   & Public Study                         & \begin{tabular}[c]{@{}c@{}} (Semi-)Structured,  \\ Markup, Text, Binary\end{tabular}                  & 15.8                  
\\ \hline
\oursys                            &  433                & 433  & 113.6                  & \makecell[c]{ Real Business,  \\ Public Dataset}                      &  \makecell[c]{ (Semi-)Structured,  \\ Markup, Text, Binary,  \\Script, Database}                 & 493.4                 
\\ \hline
\end{tabular}
}
\vspace{-2em}
\end{table}

\begin{figure}[!t]
  \centering\vspace{-.5em}
  \includegraphics[width=\linewidth]{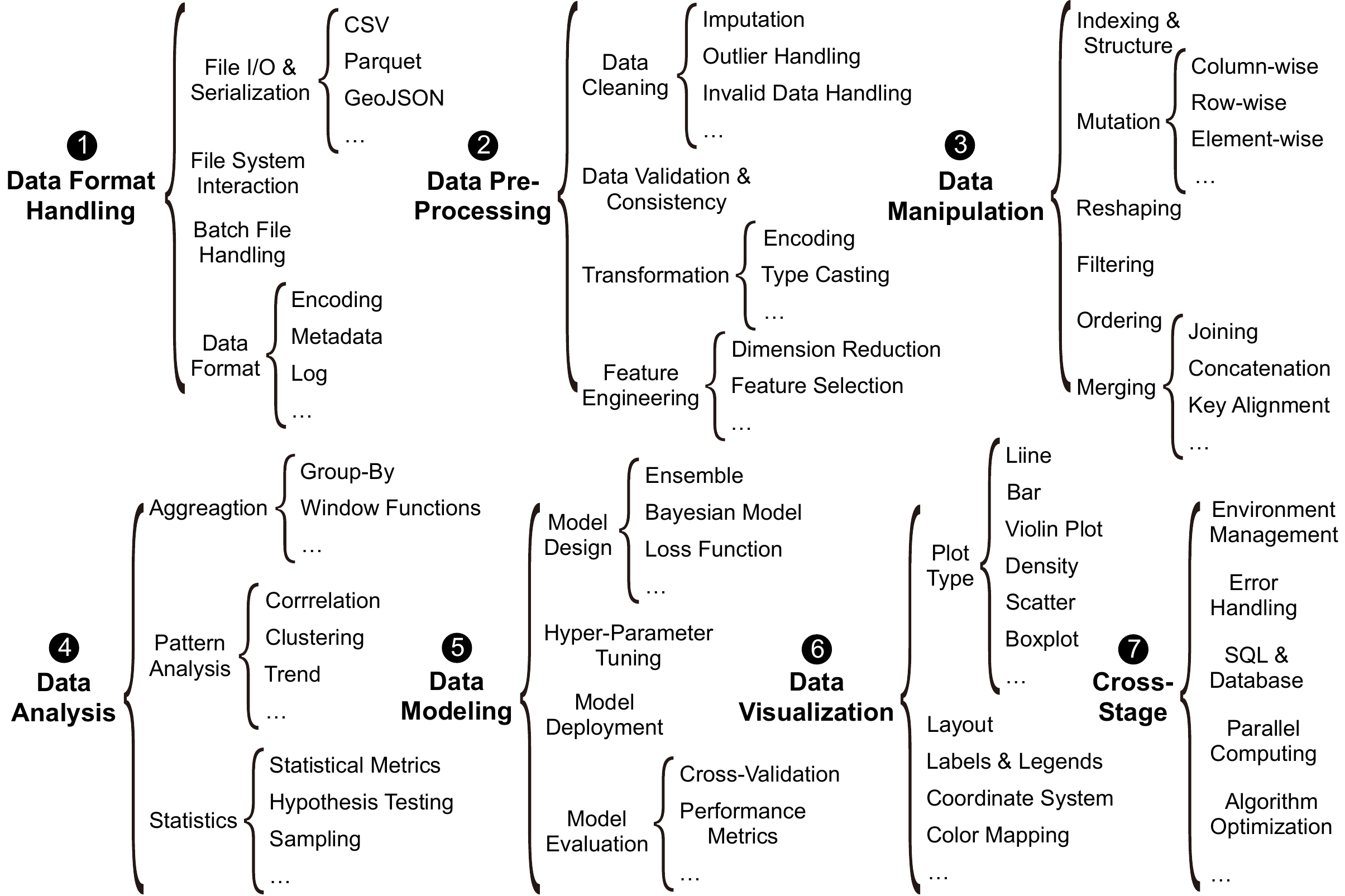}
  \vspace{-2em}
  \caption{433 Skills Generated by \oursys.}
  \label{fig:skill}
  \vspace{-1.5em}
\end{figure}

To address this gap, we propose a systematic pipeline for building a comprehensive data agent benchmark (see \autoref{fig:example}b). Our approach begins by collecting real datasets and tasks from 15 vertical domains, including 5 real-world B2B practices from a leading fintech company~\cite{antgroup}. These tasks involve complex scenarios with large-scale noisy data and long code implementations. However, these raw tasks are not directly suitable for benchmarking due to $(i)$ redundancy caused by repeated patterns with minor variations (e.g., consistently filling missing values with -1), and $(ii)$  the lack of high-quality tasks for certain domain datasets. To address this, we abstract recurring data-processing patterns shared across tasks as data science skills (e.g., ``Handling Missing Data'' in \autoref{fig:example}), and quantify benchmark coverage based on the number of skills included. From extensive task solutions, we derive a representative skill set (see \autoref{fig:skill}) and select tasks that maximize skill diversity. For generating tasks in uncovered domain datasets, we ensure benchmark quality by $(i)$  sampling realistic skill compositions, $(ii)$  promoting skill diversity across tasks, and $(iii)$  achieving comprehensive coverage of the extracted skills. Additionally, skill annotations provide the foundation for fine-grained analysis of data agent performance.

\hi{Challenges.}
There are three main challenges.
\textbf{C1: Discovery of Highly Representative Skills.}
It is non-trivial to extract data science skills from large task collections with ensured diverse representation~\cite{DBLP:journals/jmlr/ChanYYQ0022}, i.e., a relatively small set of skills that represent data science operations in solving these tasks. \textbf{C2: Task Selection for Collected Real-world Tasks.} For the collected real-world tasks, we aim to select a highly representative set of tasks that capture diverse workload patterns and scenarios while minimizing redundancy to ensure benchmarking efficiency.
\textbf{C3: Realistic Task Generation for Public Datasets.}
For public datasets that lack predefined tasks, it is essential to systematically generate realistic tasks while ensuring comprehensive coverage of underrepresented skills.

To tackle these challenges, we introduce \oursys, a comprehensive data agent benchmark built on the foundation of data-driven, discovered data science skills. First, we extract representative skills from large-scale task solutions from Stack Overflow~\cite{stackoverflow} through skill-aligned hierarchical clustering. Specifically, we leverage LLMs to break down task solutions into stepwise skill descriptions. To eliminate redundancy, we cluster semantically similar skills using pretrained text embeddings and refine each cluster through LLM-based splitting to identify distinct higher-level skills. This cluster-and-refine process is applied recursively, producing a representative set of high-level skills (addressing \textbf{C1}). Next, for real-world business tasks within each domain, we select task-solution pairs that maximize diversity in skill compositions, ensuring coverage across a wide range of practical scenarios (addressing \textbf{C2}). Finally, to ensure the benchmark comprehensively represents the extracted skills, we propose a systematic LLM-based task generation approach. This method samples frequency-aware skill compositions, uses structured dataset profiles, and generates corresponding workflows and tasks based on these skills (addressing \textbf{C3}).

\hi{Contributions.}
In summary, we make the following contributions:

\noindent (1) We propose a data science skill framework (see Section \ref{sec:pre}) and subsequently develop a comprehensive data agent benchmark, \oursys, characterized by fine-grained skill composition and real-world complexity (see Section \ref{sec:overview}).
We open-source the testbed at \blue{\url{https://github.com/AgenticDataBench/AgenticDataBench}}.

\noindent (2) We propose a hierarchical skill extraction algorithm, which performs agglomerative clustering aligned with skill boundaries using LLM-based semantic refinement (see Section \ref{sec:extract}).

\noindent (3) We propose task selection and generation modules with controlled skill coverage, including selecting skill-diverse real-world tasks and generating realistic tasks that simulate practical skill compositions (see Section~\ref{sec:benchmark}).

\noindent (4) We have conducted an in-depth fine-grained empirical study of state-of-the-art data agents, uncovering four key insights (see Section \ref{sec:experiments}).

\section{Preliminaries}\label{sec:pre}
\subsection{Data Science Benchmark}

Solving a data science task typically involves a sequence of data-related operations. We identify and summarize recurring operation patterns into skills, which represent higher-level capabilities characterized by similar application stages, technology stacks, or systematic objectives.
For example, data preprocessing can be viewed as a high-level skill that encompasses several fine-grained skills such as missing data handling and feature engineering. Together, these skills provide a multi-faceted characterization of real-world data science workflows. 

\begin{definition}[Data Science Skill]
A data science skill, $s$, is defined as a data-centric operational pattern commonly used to solve data science tasks. Formally, skills are structured in a hierarchical tree, where each skill node $s$ includes a textual description $\delta_s$ and is linked to its child skills, which represent more fine-grained capabilities. The parent-child relationships within the tree reflect abstraction and specialization among skills, with higher-level nodes representing broader, more general operations, while leaf nodes correspond to specific, actionable skill patterns.
\end{definition}

For instance, \autoref{fig:example}a presents some examples of data science skills.
``Handling Missing Data'' involves identifying \texttt{NULL} values (e.g., incomplete records or artifacts from prior processing), and applying appropriate strategies such as imputation, removal, or transformation to ensure data consistency.
We will discuss the scope of data science skills in detail in Section \ref{subsec:skill}.

Recently, data agents powered by LLMs have been introduced to automate the entire pipeline of data science tasks, from organization to execution.
To comprehensively evaluate such agents, we propose a benchmark designed to ensure broad coverage of data science skills. Each benchmark instance consists of a data science task necessitating specific skills for its resolution, along with a ground-truth solution and an evaluation function.

\begin{definition}[Data Science Benchmark Instance]
A data science benchmark instance is represented as a quintuple $(\delta_t,D,y,S,eval)$, where $\delta_t$ is a textual task objective description, $D$ is the dataset required to solve the task, $y$ is the executable task solution, $S$ is the set of skills required to solve the task as reflected in the solution $y$, and $eval$ is an evaluation function that maps the output of the data agent to a scalar score in $[0,1]$, quantifying its performance.
\end{definition}

For instance, \autoref{fig:example}a presents a representative data science benchmark. The task description specifies the required statistical computations, including rules for handling missing data and the output format (e.g., CSV). 
The dataset consists of user loan behavior records in a wide-format CSV file.
The solution is a complete implementation that satisfies both correctness and efficiency requirements.
The associated skills capture key competencies (e.g., ``Handling Missing Data'', ``Efficient Data Structures and Algorithms'') and enable fine-grained analysis of agent failures.
The evaluation function compares monthly, user-level metrics between agent outputs and the ground truth using a normalized mean squared error.
This yields a final score for ranking agents and supporting skill-level analysis (Section~\ref{subsec:pipeline}).

\subsection{Data Science Skill}
\label{subsec:skill}
In this section, we discuss the categorical scope of data science skills, and clarify how they differ from the notion of Agent Skills.

\hi{Data Science Skill Category.} Data science skills that underpin task solutions can be categorized into seven exclusive categories according to stages of the data science workflow (see \autoref{fig:skill}):
$(i)$ \emph{Data Format Handling}, including data parsing and file handling;
$(ii)$ \emph{Data Preprocessing}, including data cleaning, transformation, validation, and feature engineering;
$(iii)$ \emph{Data Manipulation}, including restructuring, indexing, filtering, modifying, and merging data;
$(iv)$ \emph{Data Analysis}, including pattern exploration, statistical computation, and aggregation for data insights;
$(v)$ \emph{Data Modeling}, including design, training, evaluation, and deployment of statistical and machine learning models;
$(vi)$ \emph{Data Visualization}, including creation of charts and other visual data representations;
$(vii)$ \emph{Cross-Stage Skills} are general-purpose skills applicable across multiple stages, such as environment management, error handling, SQL \& database, code optimization.

\hi{Discussion.} 
Some works implement agent skills as reusable modules that extend LLM capabilities~\cite{claudeskill,DBLP:conf/nips/DidolkarGKGVLRB24}, such as guidance, knowledge, scripts, and examples, dynamically incorporated to enhance scenario-specific actions. Others theoretically conceptualize skills as atomic units underlying LLM performance, analyzing outcomes at the skill level~\cite{DBLP:journals/corr/abs-2307-15936,DBLP:conf/nips/MichaudLGT23,DBLP:journals/corr/abs-2501-12391,DBLP:conf/nips/ChenRBWZSR23,DBLP:conf/iclr/00010GA25,DBLP:conf/iclr/Yu0GBGA24,DBLP:conf/iclr/MoayeriBCYFFNJV25}. Unlike these, we focus on data science scenarios, proposing a representative, data-driven skill set as a quantifiable foundation for benchmarking data agents, as elaborated in Section \ref{sec:extract}.

\begin{figure*}[!t]
  \centering
  \includegraphics[width=\linewidth]{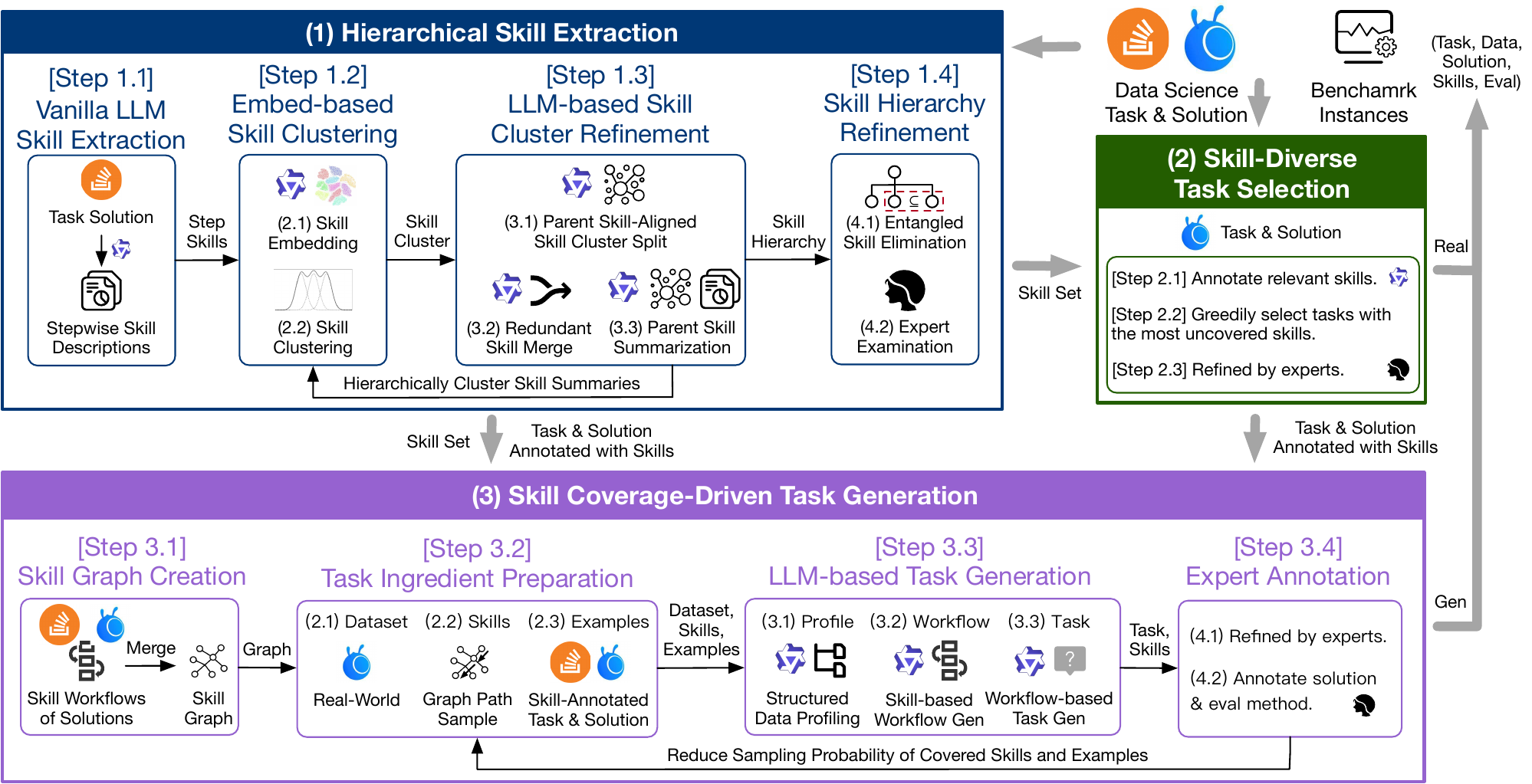}
  \vspace{-2em}
  \caption{The Workflow of Constructing \oursys.}
  \label{fig:overview}
  \vspace{-1.5em}
\end{figure*}

\section{Benchmark Overview} \label{sec:overview}

\subsection{Design Goals}
\label{subsec:goal}
We design \oursys by following the four benchmark design criteria proposed by Jim Gray ~\cite{jimmybench}.

\hi{Relevance.}
The benchmark covers a wide range of real datasets and data science task patterns.
First, we collect 97 real datasets spanning 15 domains from various sources, including 46 Kaggle datasets~\cite{kaggle}, 2 UCI ML datasets~\cite{uciml}, 2 Mendeley datasets~\cite{mendeley}, and 8 academic and government datasets (UCSD~\cite{ucsd_amazon_review}, BIRD~\cite{bird_bench}, NatEarth~\cite{naturalearthdata}, 2 from NYC TLC~\cite{nyc_tlc}, U.S. BTS~\cite{bts}, NCI GDC~\cite{nci_gdc}, OWID~\cite{owid}), and 39 real business applications at Ant Group.
Next, we abstract recurring data-centric operational patterns as data science skills, extracting 433 representative skills from 6,510 high-quality Stack Overflow data science task solutions.
Based on these skills, we generate 344 benchmark tasks that collectively cover all identified skills while simulating realistic skill compositions and usage patterns.

\hi{Simplicity.}
The benchmark is designed to reduce task redundancy and enhance clarity. 
We introduce data science skills to capture core operational patterns in task solutions, and construct the benchmark by $(i)$ selecting 102 real-world tasks from Ant Group with maximal skill diversity and $(ii)$ generating 242 additional tasks with controlled skill coverage. 
Each task and its corresponding solution are independently annotated by data science experts with the skills required, allowing for a detailed analysis of the data agent's strengths and weaknesses at the skill level.

\hi{Scalability.}
The benchmark includes large-scale datasets and diverse tasks of realistic complexity, requiring over \textbf{1,560 person-hours} of careful construction and labeling.
First, the datasets span 15 domains, totaling over 27.3 GB of data across 18 file formats, with 123.1M rows and 35.0K attributes (real business data: 5 domains, 20.1 GB, 7 file formats, 59.4M rows, 4.4K attributes).
Next, as shown in \autoref{tab:compare}, the benchmark contains 344 realistically complex data science tasks covering 433 skills, with an average of 23.5 skills per task, 113.6 lines of solution code, and 493.4 MB of data per task.

\hi{Portability.}
The benchmark is compatible with a wide range of data agent systems that accept natural language task descriptions and support data file input or manipulation.

\subsection{Benchmark Methodology Overview}

Based on these design goals, we construct the data agent benchmark using a systematic creation methodology. First, we introduce a data science skill framework to guide benchmark development, capturing recurring task patterns and enabling skill-level diversity and coverage measurement (see Section~\ref{sec:pre}). Next, we hierarchically extract representative skills from large-scale task solutions (see Section~\ref{sec:extract}). To evaluate the practical efficacy of data agents in industrial-grade scenarios, we collect real-world business datasets and tasks from a leading fintech company and reduce redundancy by selecting skill-diverse representative tasks (see Section~\ref{subsec:select}). Finally, to generate realistic tasks for other domains without predefined tasks and ensure comprehensive skill coverage, we generate tasks with realistic skill compositions that cover underrepresented areas (see Section~\ref{subsec:generate}).

\hi{Hierarchical Skill Extraction.}
A vanilla approach extracts data science skills from task solutions using LLMs by decomposing step-by-step rationales and summarizing stepwise skills~\cite{DBLP:conf/iclr/MoayeriBCYFFNJV25}.
However, such approaches often produce a large number of loosely defined skills with redundancy and entanglement.
To address this issue, we propose a hierarchical skill extraction method that further clusters related skills and abstracts higher-level skills, yielding a more compact and representative skill hierarchy.
Specifically, we first decompose task solutions into stepwise skill usage descriptions using vanilla LLM-based approach.
Next, we cluster stepwise skills using pretrained text embeddings. 
As embeddings may capture semantic details irrelevant to skill abstraction, we further prompt LLM to refine each cluster by splitting it into subclusters aligned with higher-level skill boundaries and merging redundant skills.
Then, to enhance representativeness, we recursively apply this cluster-and-refine procedure to the resulting skills to derive higher-level skills.
The process continues until the number of skills falls below a predefined threshold.
Finally, we manually refine the skill hierarchy to ensure quality.

\hi{Skill-based Benchmark Creation.}
This stage builds realistic benchmark instances across 15 domains of public datasets and real business applications, selecting skill-diverse tasks from collected data and generating realistic tasks for public datasets, collectively covering all extracted skills. Each instance comprises a task description, dataset, ground-truth solution, expert-annotated skill usage, and a task-specific evaluation method. 

\emph{\textbf{(1) Skill-Diverse Task Selection.}}
We select real-world task-solution pairs as benchmark instances by maximizing skill diversity to ensure diverse task patterns.
Specifically, given the extracted representative skills, we first prompt LLM to annotate each task with its relevant skills.
Since selecting a subset of tasks under a fixed budget to maximize skill coverage is an NP-hard problem~\cite{nemhauser1978analysis}, we adopt a greedy approximation strategy that iteratively selects the task covering the largest number of previously uncovered skills.  Finally, we manually refine the selected tasks and solutions to form complete and suitable benchmark instances.

\emph{\textbf{(2) Skill Coverage-Driven Task Generation.}}
To ensure comprehensive coverage of representative skills, we propose an LLM-based pipeline to generate practical tasks under controlled skill compositions. Specifically, we first construct a skill graph by merging skill application traces extracted from task solutions extracted from Stack Overflow and real practices, with frequency-based weights.
Next, we sample skill compositions and few-shot task–solution examples. Conditioned on them, we generate structured data profiles that model data formats and cross-dataset relationships, synthesize a skill-based workflow, and produce the corresponding task description. To encourage diversity, we apply penalties to previously covered skills and few-shot examples during the sampling process. Finally, we manually refine the task, and annotate its solution code and evaluation method to form a complete benchmark instance.

In the remainder of this section, we present the details about target test systems and evaluation pipeline of \oursys, and leave the details about hierarchical skill extraction and skill-based benchmark creation in Sections~\ref{sec:extract} and \ref{sec:benchmark}, respectively.

\begin{figure}[!t]
  \centering
  \includegraphics[width=\linewidth]{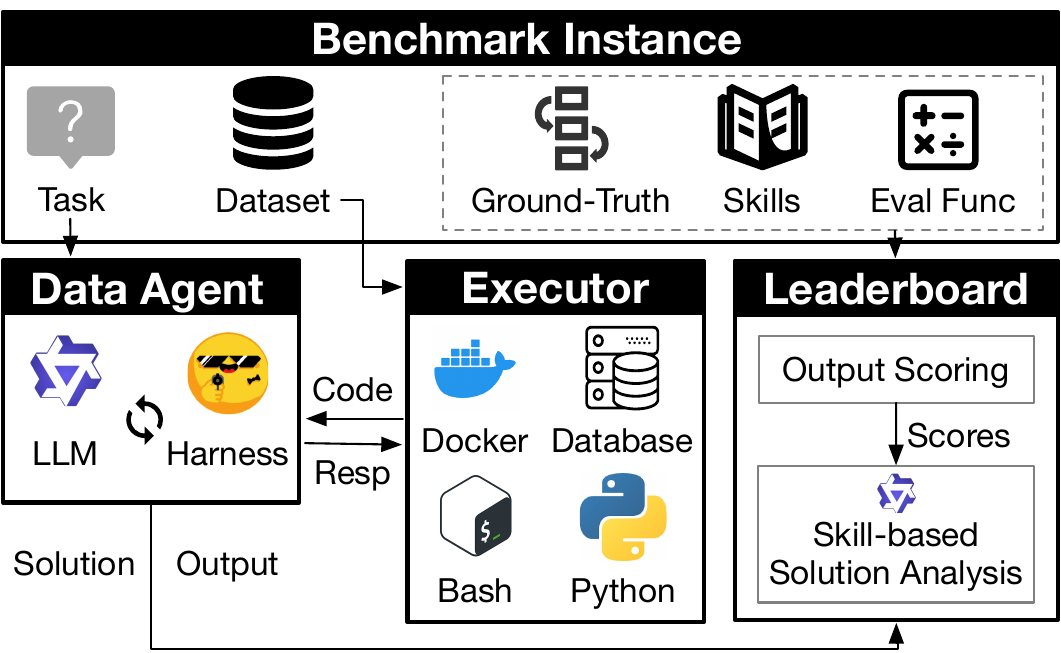}
  \vspace{-2.2em}
  \caption{The Overview of \oursys Pipeline.}
  \label{fig:testbed}
  \vspace{-2em}
\end{figure}

\subsection{\oursys Pipeline}
\label{subsec:pipeline}
As shown in \autoref{fig:testbed}, \oursys tests data agents through four components: \textit{Benchmark Instances}, \textit{Data Agent}, \textit{Executor}, and \textit{Leaderboard}. First, we load \textit{Benchmark Instances}, each of which includes a data science task, dataset, the ground-truth solution and answer, required skills, and an evaluation function for scoring task performance. Second, the \textit{Executor} is prepared within a Docker image, supporting Bash, Python, and database operations, and the benchmark dataset is loaded for exploration and execution.
Third, the task is passed to the LLM-driven \textit{Data Agent}, which iteratively generates and executes code in the \textit{Executor} based on previous execution responses, and produces a final solution and output. The \textit{Leaderboard} evaluates the solutions generated by the data agent, compares them against the ground truth, and ranks the data agent on the leaderboard based on its evaluation score. The evaluation process is conducted in two steps. (1) The evaluation function produces a performance score. \oursys supports five scoring modes:
$(i)$ table matching, which checks equality between the predicted and ground-truth tables, allowing tolerance thresholds for numerical columns and exact matching for others;
$(ii)$ modeling-based scoring, e.g., mean squared error, normalized to $[0,1]$; 
$(iii)$ JSON matching, i.e., the proportion of keys with matched values (approximate matching for numerical fields within thresholds, exact matching otherwise);
$(iv)$ chart matching, which compares both underlying numerical data and plot configurations;
$(v)$ exact and fuzzy text matching.
(2) We conduct skill-level analysis by prompting LLMs to compare the solution with the ground truth, using annotated skills as candidates and scoring information to identify incorrectly applied skills as root causes of performance gaps.

\section{Hierarchical Skill Extraction}
\label{sec:extract}

We discuss how to extract a representative set of data science skills from a large corpus of data science task solutions.
These skills comprehensively capture recurring data-related operational patterns across the solutions. 
Building on this skill framework, we construct benchmark data science tasks with controlled coverage and divergence among selected and newly generated tasks.

\hi{Step 1: Vanilla LLM-based Skill Extraction.}
We first collect 6,510 data science tasks and solutions from Stack Overflow~\cite{stackoverflow}, filtering by $(i)$ relevant tags such as ``data-science'' and ``data-analysis,'' and $(ii)$ quality indicators, including accepted answers or scores higher than 3.
Next, since many solutions involve complex pipelines that require multiple data science skills, we employ LLM to decompose each solution into stepwise rationales of skill usage~\cite{DBLP:conf/iclr/MoayeriBCYFFNJV25}.
Specifically, we prompt LLM to ensure that each step corresponds to a distinct data science skill while preserving actionable details, and that the collection of steps collectively reconstructs the original solution.
In total, this process yields 29,602 stepwise skill descriptions.

However, these skills are unsuitable as a basis for benchmark creation due to three main drawbacks.
$(i)$ \textit{Scalability.}
Tens of thousands of loosely defined skills hinder the construction of an efficient and effective benchmark at a manageable scale (see \autoref{tab:compare}).
$(ii)$ \textit{Redundancy.}
A common issue is that many skill descriptions refer to the same underlying skill, reducing the diversity of the skill set.
$(iii)$ \textit{Entanglement.}
We also observe a prevalent entanglement phenomenon among extracted skills, where one skill represents a high-level abstraction that subsumes another. Such skills should not be simultaneously retained as representative.

\hi{Step 2: Embedding-based Skill Clustering.}
The above drawbacks can be mitigated by adopting hierarchical clustering, where each cluster corresponds to a smaller set of higher-level skill abstractions. Specifically, during the agglomerative clustering process~\cite{murtagh2012algorithms}, $(i)$ the number of top-level skills naturally decreases as clusters are progressively merged, and $(ii)$ redundant or entangled skills, which often share similar semantics, are prone to being merged.

First, we adopt a state-of-the-art text embedding model Qwen3-Embedding~\cite{qwen3embed} to encode skill descriptions into vectors.
We also apply UMAP~\cite{mcinnes2018umap} to reduce the embedding dimension while preserving local data manifold structures.
Next, we apply GMM~\cite{sarthiraptor} for soft clustering, allowing skills to be associated with multiple higher-level skills.
Since embedding vector similarities may fail to capture shared high-level skill abstractions due to irrelevant details (e.g., formats or topics), we further use LLMs to split each cluster and align it with coherent higher-level skills (see Step 3).
To fit LLM context, we constrain the number of skill descriptions in each cluster below a predefined threshold.
If a cluster exceeds this threshold, we recursively apply GMM to split it into smaller clusters.

\hi{Step 3: LLM-based Skill Cluster Refinement.}
We split each cluster into sub-clusters representing higher-level skill abstractions, yielding a more compact hierarchy than the original low-level skill annotations. Specifically, for each cluster of semantically related skills, we prompt LLMs to derive higher-level skills and group the low-level skills accordingly. We also preserve the lineage between low-level skills and their parent skills.

Since many skills correspond to synonymous skills, we apply DBSCAN~\cite{ester1996density} to detect and merge them. 
Specifically, we embed the LLM-generated skill descriptions using Qwen3-Embedding model, and perform clustering with a strict distance threshold to separate semantically divergent skills. We then merge the synonymous skills, and select the shortest skill description as their representative.

If the resulting top-level skills remain too numerous for a manageable benchmark scale, we further repeat the cluster-and-refine process to derive fewer higher-level skills.
Specifically, we generate a summary for each skill by augmenting the LLM-generated description with representative solution steps that exhibit the largest average cosine similarity to the other steps using the skill.
We recursively cluster these skill summaries until the number of top-level skills falls below a predefined threshold.

\hi{Step 4: Skill Hierarchy Refinement.}
We further address entanglement in the skill hierarchy, where an LLM-generated skill is often overly general if it subsumes another skill at the same or a shallower level.
Specifically, we extract syntactic tokens from skill descriptions and identify entanglement via token-set subset relations, assuming such containment indicates semantic subsumption.
We then replace overly general skills with their more specific children, and update the hierarchy accordingly.

Finally, we engage data science experts to review the top-level skills to ensure they are appropriately scoped, diverse, representative of common data-related operations in practice, and aligned with realistic evaluation scenarios.
Through this process, we obtain 433 top-level skills.

\section{Skill-based Benchmark Creation}
\label{sec:benchmark}

We describe the construction of \oursys based on the extracted data science skill set.
Each benchmark instance consists of a task description grounded in real-world datasets, a ground-truth solution, annotated skills required to solve the task, and a task-specific evaluation method. To reflect practical data science challenges, we incorporate real-world datasets and tasks across real business domains within a leading fintech company (see Section~\ref{subsec:select}). Then, to ensure cross-domain coverage and comprehensive skill coverage, we generate additional tasks over real-world datasets from diverse domains with specific skill compositions (see Section~\ref{subsec:generate}).

\subsection{Skill-Diverse Task Selection}
\label{subsec:select}

Given massive corpus of anonymized production data from Ant Group's B2B ecosystem, 30 domain experts from 5 business units spent over \textbf{600 person-hours} curating 600 representative and complex tasks from real-world practice.
These tasks span diverse industries, including commercial banking, consumer finance, internet finance, insurance, automotive, aviation, mobile manufacturing, and retail.
They also cover a wide range of scenarios (e.g., exploratory analysis, modeling, operations), and preserve realistic challenges such as noise, long-tailed distributions, and feature leakage.

Since many tasks exhibit similar operational structures with variations only in parameters or datasets, we propose a skill-diverse task selection method that adopts the skill framework to maximize coverage across diverse task patterns while reducing redundancy.
We implement the method in three steps:

\hi{Step 1: Relevant Skill Annotation.}
For each task, we use LLMs to annotate relevant skills.
Specifically, we evaluate the presence of each candidate skill in the task solution independently using the asynchronous batch inference mode of the Bailian platform~\cite{bailian}.
Next, we input the solution and the identified skills into LLMs and prompt LLMs to infer skill dependencies and generate a skill usage trace, which is later used to simulate skill composition during task generation (see Section~\ref{subsec:generate}).

\hi{Step 2: Skill-Diverse Task Selection.}
To enhance benchmark efficiency, we select a representative subset of tasks under a predefined budget while maximizing skill coverage.
This can be formulated as an NP-hard problem that maximizes a submodular set function, i.e., the number of skills covered by the selected task set.
It admits a $1-1/e$ approximation guarantee via a greedy algorithm~\cite{nemhauser1978analysis}.
Specifically, we iteratively select tasks, each time choosing the task that covers the largest number of previously uncovered skills, until the selected task set covers all candidate skills.

\hi{Step 3: Expert Refinement.}
To curate tasks suitable for benchmarking, human experts design task-specific evaluation functions, review datasets to ensure the absence of privacy concerns, and refine task descriptions and skill annotations.
Though this process, we obtain 102 benchmark instances from real-world business practices.

\subsection{Skill Coverage-Driven Task Generation}
\label{subsec:generate}

To enhance benchmark coverage, we further incorporate 58 datasets from popular open repositories spanning 10 previously uncovered domains, including 46 Kaggle datasets, 2 UCI ML datasets, 2 Mendeley datasets, and 8 academic and government datasets (UCSD, BIRD, NatEarth, 2 from NYC TLC, U.S. BTS, NCI GDC, OWID).
We select these repositories based on three criteria: $(i)$ real-world relevance to prevalent data science domains; $(ii)$ inherent complexity, including large-scale data, complex file structures, noisy content, and heterogeneous formats; and $(iii)$ flexible cross-file associations, such as overlapping semantic topics or joinable attributes (e.g., time, users, countries).

We generate tasks with realistic skill compositions absent from the collected tasks.
Specifically, we design a skill coverage-driven task generation method.
First, we construct a skill graph by merging skill application traces from task solutions collected from Stack Overflow and real practices.
The node and edge weights reflect real-world frequencies of skills and their dependencies.
Next, to generate a practical task, we prepare key ingredients including real-world datasets, sampled paths from the skill graph, and skill-relevant tasks and solutions as references.
Then, we employ a systematic LLM-based pipeline to generate the task, including structured data profiling, workflow synthesis using sampled skills, and task construction grounded in the workflow with quality verification.
To promote benchmark diversity, we also dynamically reduce sampling weights of previously covered skills and reference examples.
Finally, human experts refine the generated tasks to ensure alignment with the skills, and curate corresponding solutions and evaluation methods to produce complete benchmark instances.

\hi{Step 1: Skill Graph Creation.}
To enhance realism, we simulate real skill compositions by building a skill graph based on aggregated skill usage traces from task solutions. Specifically, for Stack Overflow tasks, we trace the extracted skills back to their original solution steps, and derive skill traces from the step sequences within each solution (see Section \ref{sec:extract}). For real business tasks from Ant Group, we directly utilize the skill traces obtained in Step 1 of Section \ref{subsec:select}.
Based on these skill-annotated task solutions, we construct a skill graph where nodes denote skills and edges represent consecutive skill usage in task solutions.
Node and edge weights are frequencies of individual skills and ordered skill pairs, respectively.

\hi{Step 2: Task Ingredient Preparation.}
Before creating a new task, we prepare three necessary ingredients:

\emph{\textbf{$(1)$ Dataset.}}
We load datasets from the target domain.

\emph{\textbf{$(2)$ Skills.}}
We sample a skill composition by drawing a random path from the skill graph.
Specifically, the starting node is sampled from skills that appear within the first 10\% of steps in some task solutions, with probabilities proportional to node weights.
Each subsequent node is sampled from the neighbors of the current node, with probability proportional to a weighted combination of the corresponding edge weight and neighbor node weight.
Sampling continues until the path reaches the predefined length.

\emph{\textbf{$(3)$ Examples.}}
We retrieve representative task-solution pairs relevant to the sampled skills to guide task generation.
Specifically, first, for each skill, we assign a relevance score to its annotated steps, defined as the average cosine similarity between the step and other steps annotated with the same skill, plus one to ensure non-negativity.
Steps not associated with the skill receive a relevance score of zero.
Then, given the sampled skills, we compute the relevance score of each task–solution pair by summing the step-wise relevance scores for each skill and aggregating them across all sampled skills.
Finally, we sample a predefined number of tasks with probabilities proportional to their aggregated relevance scores, and retain them as examples of skill application.

\hi{Step 3: LLM-based Task Generation.}
Leveraging the prepared dataset, sampled skills, and skill-related examples, we use LLM to generate new tasks through three stages:

\emph{\textbf{$(1)$ Structured Data Profiling.}}
We create a data profile for each dataset file to provide structured information to LLMs, consisting of three components:
$(i)$ \textit{Basic Information}, which applies to all data files and includes the file path, number of rows, and sampled initial rows;
$(ii)$ \textit{Data Format-Specific Structure}, represented as a structured dictionary where each key describes a key attribute of the data file.
For example, for tabular data, it includes columns, column types, numerical and categorical columns, missing values, delimiters, and detected header rows determined by textual value ratios or LLM-based distinction between metadata and tabular content;
$(iii)$ \textit{Relationship}, which captures potential join relationships among attributes across data files.
Specifically, we first programmatically identify attribute names shared across files within the same subfolder.
To capture subtler semantic relationships, we prompt LLMs with the data format-specific structure of each file to generate cross-source relationships including fuzzy attribute matches, thematic parallels, and suggested joins.
All discovered joins are manually reviewed to ensure the correctness.

\emph{\textbf{$(2)$ Skill-based Workflow Generation.}}
Since the generated task should require the sampled skills for solution, we first synthesize a solution workflow based on these skills and then generate the task accordingly.
$(i)$ \textit{Initialization.}
We first sample one skill and its examples, and prompt LLM to generate an actionable step that potentially involves multiple correlated data files conditioned on data profiles.
If there are too many data files, we cluster files with similar name patterns (e.g., differing only by indices) or tables within the same directory that share schema.
We then provide the clustered file paths and a predefined number of representative data profiles to LLM.
$(ii)$ \textit{Skill Iteration.}
For each remaining sampled skill, we iteratively insert it into the workflow by prompting the LLM to generate a step using the skill, determine its position in the workflow, and update step dependencies accordingly.
After insertion, we require LLM to verify step actionability and dependency coherence.
If verification fails, we retry until a predefined failure threshold is reached.
If the threshold is exceeded, the skill is discarded.
$(iii)$ \textit{Termination.}
We repeat this process until all skills are either integrated into or excluded from the workflow.

\emph{\textbf{$(3)$ Workflow-based Task Generation.}}
Based on the workflow steps annotated with used skills and data files, we prompt LLM to generate a task description with verification to ensure six quality criteria, including solvability by the workflow, necessity of the skills, conciseness, clarity, actionability, and a verifiable answer.
We retain only tasks that pass these verifications.

\emph{\textbf{$(4)$ Dynamic Sampling Penalty.}}
To enhance diversity among the generated tasks, we first penalize repeated skills by dividing the weights of previously covered nodes and edges by one plus their sampling count. 
We also apply the same penalization to the relevance scores of previously used task–solution pairs.

\hi{Step 4: Expert Annotation.}
we establish a systematic annotation pipeline to ensure the quality of generated benchmark instances, including: $(i)$ validating pipeline correctness; $(ii)$ identifying missing or redundant data sources in each step; $(iii)$ refining questions to better evaluate skill application; $(iv)$ assessing question quality in terms of conciseness, clarity, and domain relevance; $(v)$ designing evaluation functions; and $(vi)$ implementing ground-truth solutions. We further conduct multiple rounds of cross-validation to ensure annotation consistency and reliability. This pipeline engages 8 experts and requires \textbf{960 person-hours}, resulting in 242 benchmark instances derived from real-world public datasets.

\section{Experiments}
\label{sec:experiments}

\subsection{Experimental Setup}
\label{subsec:setup}
All experiments are conducted on a Linux server with 128 GB RAM and a 3.1 GHz CPU.
We execute data agents in a Docker environment to ensure safety and consistent evaluation.

\begin{table}[!t]
\caption{Statistics Across 15 \oursys Domains.}
\vspace{-1em}
\label{tab:dataset}
\resizebox{\linewidth}{!}{
\begin{tabular}{|cc|c|c|ccc|}
\hline
\multicolumn{2}{|c|}{\multirow{2}{*}{\textbf{Domain}}}                                                                     & \multirow{2}{*}{\textbf{\# Files}} & \multirow{2}{*}{\begin{tabular}[c]{@{}c@{}}\textbf{Data}\\ \textbf{(GB)}\end{tabular}} & \multicolumn{3}{c|}{\textbf{Per Task}} \\ \cline{5-7} 
\multicolumn{2}{|c|}{}                                                                                            &                           &                                                                      & \multicolumn{1}{c|}{\textbf{Files}} & \multicolumn{1}{c|}{\begin{tabular}[c]{@{}c@{}}\textbf{Data} \\ \textbf{(MB)}\end{tabular}} & \textbf{\# Skills} \\ \hline
\multicolumn{1}{|c|}{\multirow{5}{*}{\begin{tabular}[c]{@{}c@{}}Real\\Business\end{tabular}}} & Financial
& \multicolumn{1}{c|}{6}
& 0.1
& \multicolumn{1}{c|}{ 6.0}
& \multicolumn{1}{c|}{ 89.4}
&  14.3 \\ \cline{2-7}
\multicolumn{1}{|c|}{} & Loan Model
& \multicolumn{1}{c|}{121}
& 0.03
& \multicolumn{1}{c|}{ 33.7}
& \multicolumn{1}{c|}{ 24.6}
&  20.8 \\ \cline{2-7}
\multicolumn{1}{|c|}{} & Loan Risk
& \multicolumn{1}{c|}{39}
& 16.1
& \multicolumn{1}{c|}{ 1.1}
& \multicolumn{1}{c|}{ 397.5}
&  16.4 \\ \cline{2-7}
\multicolumn{1}{|c|}{} & Marketing
& \multicolumn{1}{c|}{4}
& 3.3
& \multicolumn{1}{c|}{ 4.0}
& \multicolumn{1}{c|}{ 3304.1}
&  18.1 \\ \cline{2-7}
\multicolumn{1}{|c|}{} & Strategy
& \multicolumn{1}{c|}{4}
& 0.5
& \multicolumn{1}{c|}{ 1.0}
& \multicolumn{1}{c|}{ 207.7}
&  12.9 \\ \hline
\multicolumn{1}{|c|}{\multirow{10}{*}{\begin{tabular}[c]{@{}c@{}}Public\\Dataset\end{tabular}}} & Agriculture
& \multicolumn{1}{c|}{14}
& 0.2
& \multicolumn{1}{c|}{ 3.7}
& \multicolumn{1}{c|}{ 58.1}
&  27.2 \\ \cline{2-7}
\multicolumn{1}{|c|}{} & Ecommerce
& \multicolumn{1}{c|}{12}
& 3.1
& \multicolumn{1}{c|}{ 7.2}
& \multicolumn{1}{c|}{ 2602.4}
&  27.7 \\ \cline{2-7}
\multicolumn{1}{|c|}{} & Energy
& \multicolumn{1}{c|}{9}
& 0.1
& \multicolumn{1}{c|}{ 5.3}
& \multicolumn{1}{c|}{ 73.8}
&  26.7 \\ \cline{2-7}
\multicolumn{1}{|c|}{} & Entertainment
& \multicolumn{1}{c|}{14}
& 1.0
& \multicolumn{1}{c|}{ 6.0}
& \multicolumn{1}{c|}{ 93.7}
&  22.3 \\ \cline{2-7}
\multicolumn{1}{|c|}{} & Healthcare
& \multicolumn{1}{c|}{15}
& 0.2
& \multicolumn{1}{c|}{ 5.4}
& \multicolumn{1}{c|}{ 38.9}
&  29.5 \\ \cline{2-7}
\multicolumn{1}{|c|}{} & Real Estate
& \multicolumn{1}{c|}{48}
& 0.4
& \multicolumn{1}{c|}{ 7.5}
& \multicolumn{1}{c|}{ 64.1}
&  24.8 \\ \cline{2-7}
\multicolumn{1}{|c|}{} & Sports
& \multicolumn{1}{c|}{18}
& 0.4
& \multicolumn{1}{c|}{ 4.0}
& \multicolumn{1}{c|}{ 231.0}
&  28.8 \\ \cline{2-7}
\multicolumn{1}{|c|}{} & Social Network
& \multicolumn{1}{c|}{8}
& 0.5
& \multicolumn{1}{c|}{ 3.2}
& \multicolumn{1}{c|}{ 27.3}
&  24.4 \\ \cline{2-7}
\multicolumn{1}{|c|}{} & Tourism
& \multicolumn{1}{c|}{18}
& 0.04
& \multicolumn{1}{c|}{ 6.4}
& \multicolumn{1}{c|}{ 37.6}
&  24.5 \\ \cline{2-7}
\multicolumn{1}{|c|}{} & Transportation
& \multicolumn{1}{c|}{12}
& 1.3
& \multicolumn{1}{c|}{ 8.9}
& \multicolumn{1}{c|}{ 980.2}
&  27.1 \\ \hline

\multicolumn{2}{|c|}{Total}
& \multicolumn{1}{c|}{342}
&  27.3
& \multicolumn{1}{c|}{ 6.4}
& \multicolumn{1}{c|}{ 493.4}
&  23.5 \\
\hline
\end{tabular}
}
\vspace{-1.5em}
\end{table}

\begin{table*}[!t]
\centering
\caption{Representative (TF-IDF) and Challenging (Score) Skills by Domain (DFH: Data Format Handling, DP: Data Preprocessing, DM: Data Manipulation, DA: Data Analysis, DML: Data Modeling, DV: Data Visualization, CS: Cross-Stage Skills). TF-IDF ranks skills by frequency scaled by $log($total tasks$/$skill tasks$)$ (the higher, the more frequent); challenging skills are those with the lowest aggregated LLM-assigned scores across domain-relevant tasks (the lower, the more challenging).}
\vspace{-1em}
\label{tab:domain-skill}
\resizebox{\linewidth}{!}{
\begin{tabular}{|c|c|c|c|c|}
\hline
\textbf{Domain} & \textbf{Representative Skill (TF-IDF)} & \textbf{Category} & \textbf{Challenging Skill (Score)} & \textbf{Category} \\
\hline
\multirow{2}{*}{Financial} & Metadata and Documentation Review (64.48) & DFH & SQL Optimization and Advanced Usage (0.50) & CS \\ \cline{2-5}
 & Query Construction and Execution (34.25) & CS & Data Transformation and Calculation (0.58) & DM \\ \hline
\multirow{2}{*}{Loan Model} & DataFrame Column Management (6.59) & DM & Data Comparison and Validation (0.32) & DA \\ \cline{2-5}
 & Model Training and Customization (4.40) & DML & Statistical Testing for Feature-Target Evaluation (0.34) & DA \\ \hline
\multirow{2}{*}{Loan Risk} & Custom Value Replacement and Correction (6.44) & DP & Data Preprocessing and Column Management (0.30) & DP \\ \cline{2-5}
 & Helper Functions and Reusable Code (4.83) & CS & Normalization and Percentile Calculations (0.34) & DA \\ \hline
\multirow{2}{*}{Marketing} & Model Training and Customization (6.29) & DML & Performance Metrics and Optimization (0.27) & CS \\ \cline{2-5}
 & Performance Benchmarking and Evaluation (4.83) & CS & Computational Frameworks and Libraries (0.30) & CS \\ \hline
\multirow{2}{*}{Strategy} & Event Tracking and Funnel Analysis (4.03) & DA & Data Preprocessing and Column Management (0.13) & DP \\ \cline{2-5}
 & Data Preprocessing and Segmentation (2.20) & DP & Data Preprocessing and Segmentation (0.20) & DP \\ \hline
\multirow{2}{*}{Agriculture} & Entity Mapping and Matching (6.44) & DA & Probability Modeling and Conversion (0.24) & DA \\ \cline{2-5}
 & Data Exploration and Comparison (5.33) & DA & Time Series Analysis and Causality (0.25) & DA \\ \hline
\multirow{2}{*}{Ecommerce} & Downsampling and Resampling (8.06) & DA & Statistical Analysis and Testing (0.11) & DA \\ \cline{2-5}
 & String and Categorical Data Handling (8.05) & DM & Data Analysis and Visualization (0.23) & DV \\ \hline
\multirow{2}{*}{Energy} & Mapping and Lookup (3.81) & DM & Normalization and Percentile Calculations (0.28) & DA \\ \cline{2-5}
 & Reshaping and Aggregation (3.22) & DA & Statistical Modeling and Uncertainty (0.33) & DA \\ \hline
\multirow{2}{*}{Entertainment} & Data Extraction from JSON (3.97) & DFH & Data Categorization \& Mapping (0.27) & DM \\ \cline{2-5}
 & Data Structure and Dictionary Operations (2.64) & DA & Regression Modeling and Interpretation (0.27) & DML \\ \hline
\multirow{2}{*}{Healthcare} & Special Data Handling and Padding (12.09) & DM & Incremental and Comparative Calculations (0.25) & DA \\ \cline{2-5}
 & ETL and Data Integration (6.91) & DP & Time Series and Window Analysis (0.28) & DA \\ \hline
\multirow{2}{*}{Real Estate} & DataFrame Transformation and Reshaping (9.66) & DM & Correlation Matrix Generation (0.17) & DA \\ \cline{2-5}
 & Date Adjustment and Alignment (6.10) & DM & Data Manipulation and Validation (0.23) & DM \\ \hline
\multirow{2}{*}{Sports} & ETL and Data Integration (10.69) & DP & Rolling Statistics and Window-Based Signal Processing (0.20) & DA \\ \cline{2-5}
 & Data Alignment \& Merging (7.15) & DM & Gradient and Derivative Methods (0.26) & DM \\ \hline
\multirow{2}{*}{Social Network} & Encoding and Format Identification (8.79) & DFH & Mathematical Foundations and Algorithm Understanding (0.20) & CS \\ \cline{2-5}
 & ETL and Data Integration (8.17) & DP & Validation and Verification of Merge Results (0.27) & DA \\ \hline
\multirow{2}{*}{Tourism} & Excel File Handling and Automation (45.06) & DFH & Ranking and Top N Logic (0.20) & DM \\ \cline{2-5}
 & Command-Line and Shell Operations (9.30) & CS & Ranking and Normalization (0.27) & DM \\ \hline
\multirow{2}{*}{Transportation} & Compression and Archiving (62.46) & DFH & Time Series Analysis and Causality (0.18) & DA \\ \cline{2-5}
 & Time Series Alignment and Matching (36.27) & DM & Topic Modeling and Evaluation (0.19) & DML \\ \hline
\end{tabular}
}
\vspace{-1em}
\end{table*}

\begin{figure*}[!t]
  \centering
  \includegraphics[width=\linewidth]{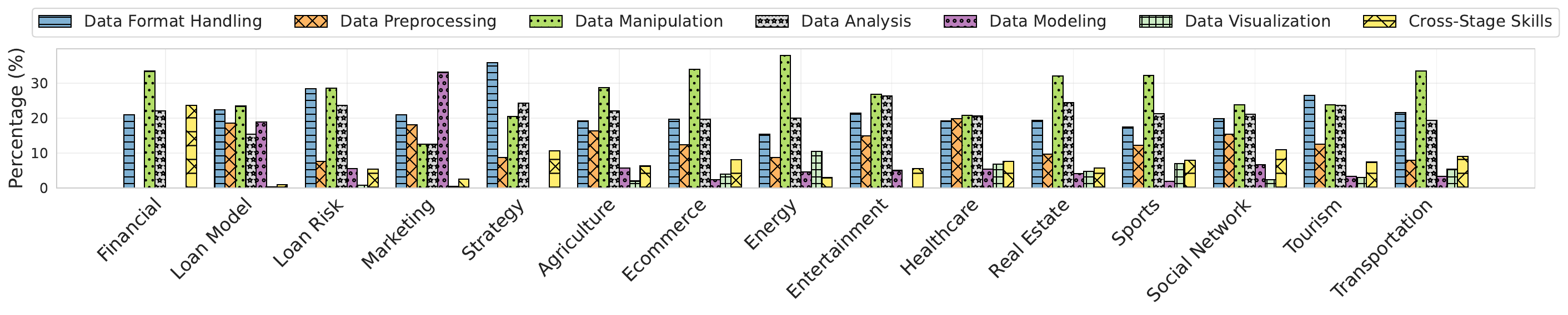}
  \vspace{-2.5em}
  \caption{Skill Category Distribution across Domain.}
  \vspace{-1.5em}
  \label{fig:domain-skill}
\end{figure*}

\begin{table*}[!t]\vspace{-1em}
\caption{Scores (\%) over \oursys. SA=Smolagents, DA=DA-Agent, CC=Claude Code, CX=CodeX. \ding{192}=Qwen3.5-397B-A17B, \ding{193}=Kimi-K2.5, \ding{194}=Claude Sonnet 4.6.}
\label{tab:overall}
\vspace{-1em}
\resizebox{\linewidth}{!}{
\begin{tabular}{|cc|c|c|c|c|c|c|c|c|c|c|c|c|}
\hline
\multicolumn{2}{|c|}{\textbf{Domain}} & \begin{tabular}[c]{@{}c@{}}\emph{SA}\\ \emph{(\ding{192})}\end{tabular} & \begin{tabular}[c]{@{}c@{}}\emph{SA}\\ \emph{(\ding{193})}\end{tabular} & \begin{tabular}[c]{@{}c@{}}\emph{SA}\\ \emph{(\ding{194})}\end{tabular} & \begin{tabular}[c]{@{}c@{}}\emph{DA}\\ \emph{(\ding{192})}\end{tabular} & \begin{tabular}[c]{@{}c@{}}\emph{DA}\\ \emph{(\ding{193})}\end{tabular} & \begin{tabular}[c]{@{}c@{}}\emph{DA}\\ \emph{(\ding{194})}\end{tabular} & \begin{tabular}[c]{@{}c@{}}\emph{CC}\\ \emph{(\ding{192})}\end{tabular} & \begin{tabular}[c]{@{}c@{}}\emph{CC}\\ \emph{(\ding{193})}\end{tabular} & \begin{tabular}[c]{@{}c@{}}\emph{CC}\\ \emph{(\ding{194})}\end{tabular} & \begin{tabular}[c]{@{}c@{}}\emph{CX}\\ \emph{(\ding{192})}\end{tabular} & \begin{tabular}[c]{@{}c@{}}\emph{CX}\\ \emph{(\ding{193})}\end{tabular} & \begin{tabular}[c]{@{}c@{}}\emph{CX}\\ \emph{(\ding{194})}\end{tabular} \\ \hline
\multicolumn{1}{|c|}{\multirow{5}{*}{\begin{tabular}[c]{@{}c@{}}Real\\ Business\end{tabular}}} & Financial & 58.1 & 61.5 & 54.9 & 54.9 & 65.4 & 58.6 & 55.5 & 64.3 & 59.1 & 60.7 & \textbf{66.9} & 52.9 \\ \cline{2-14}
\multicolumn{1}{|c|}{} & Loan Model & 41.1 & 42.3 & 41.2 & \textbf{43.9} & 43.4 & 41.6 & 41.5 & 42.9 & 42.5 & 33.6 & 40.6 & 39.7 \\ \cline{2-14}
\multicolumn{1}{|c|}{} & Loan Risk & 72.0 & 72.9 & 69.0 & 70.9 & 69.7 & 71.6 & 74.0 & 72.7 & 72.8 & 58.9 & \textbf{74.0} & 52.7 \\ \cline{2-14}
\multicolumn{1}{|c|}{} & Marketing & 37.3 & 43.5 & \textbf{47.5} & 36.8 & 33.4 & 39.4 & 31.9 & 35.4 & 46.2 & 29.6 & 32.2 & 18.9 \\ \cline{2-14}
\multicolumn{1}{|c|}{} & Strategy & 43.1 & 45.4 & 45.4 & 36.3 & 47.4 & \textbf{50.8} & 42.9 & 39.2 & 48.0 & 38.2 & 44.8 & 34.1 \\ \hline
\multicolumn{1}{|c|}{\multirow{10}{*}{\begin{tabular}[c]{@{}c@{}}Public\\ Dataset\end{tabular}}} & Agriculture & 37.9 & 32.6 & 40.3 & 33.3 & 35.7 & 31.3 & 37.1 & 38.5 & 37.1 & 32.9 & \textbf{40.9} & 21.2 \\ \cline{2-14}
\multicolumn{1}{|c|}{} & Ecommerce & 31.8 & 27.9 & 35.8 & 29.9 & 28.0 & 30.4 & 31.0 & 30.6 & 30.0 & 26.5 & \textbf{36.3} & 18.1 \\ \cline{2-14}
\multicolumn{1}{|c|}{} & Energy & 63.1 & 48.9 & 61.6 & 59.0 & 41.3 & 58.9 & 54.7 & 47.1 & 59.3 & 51.2 & \textbf{69.0} & 46.8 \\ \cline{2-14}
\multicolumn{1}{|c|}{} & Entertainment & \textbf{69.3} & 58.9 & 51.1 & 63.3 & 57.2 & 60.3 & 46.4 & 52.0 & 57.7 & 29.0 & 50.7 & 42.8 \\ \cline{2-14}
\multicolumn{1}{|c|}{} & Healthcare & \textbf{33.3} & 32.6 & 29.0 & 26.5 & 26.4 & 25.4 & 32.7 & 27.8 & 31.4 & 28.3 & 29.9 & 14.0 \\ \cline{2-14}
\multicolumn{1}{|c|}{} & Real Estate & 22.4 & 22.9 & 22.9 & 19.3 & \textbf{25.0} & 17.8 & 14.1 & 21.0 & 21.7 & 18.1 & 23.3 & 8.1 \\ \cline{2-14}
\multicolumn{1}{|c|}{} & Sports & 40.6 & 34.0 & 39.4 & 37.1 & 39.5 & 43.9 & 29.0 & 37.6 & 42.0 & 35.4 & \textbf{44.0} & 25.7 \\ \cline{2-14}
\multicolumn{1}{|c|}{} & Social Network & 59.7 & 48.6 & 67.0 & 58.8 & 60.1 & \textbf{68.0} & 45.1 & 55.3 & 60.9 & 59.7 & 63.3 & 33.1 \\ \cline{2-14}
\multicolumn{1}{|c|}{} & Tourism & 56.6 & 53.8 & 55.9 & 57.2 & 55.4 & \textbf{60.4} & 49.9 & 49.2 & 54.5 & 46.5 & 56.8 & 42.0 \\ \cline{2-14}
\multicolumn{1}{|c|}{} & Transportation & 49.9 & 37.4 & 46.2 & 45.8 & 39.7 & 44.8 & 32.7 & 42.6 & 45.1 & 36.3 & \textbf{53.4} & 34.5 \\ \hline
\multicolumn{2}{|c|}{Total} & 47.1 & 43.8 & 46.7 & 44.4 & 44.8 & 46.1 & 40.9 & 44.3 & 46.6 & 39.9 & \textbf{48.8} & 31.6 \\ \hline
\end{tabular}
}
\vspace{-1em}
\end{table*}

\hi{Evaluated Methods.}
We evaluate state-of-the-art LLMs, including open-source Qwen3.5-397B-A17B, Kimi-K2.5, and the closed-source Claude Sonnet 4.6.
We use default temperatures.
We evaluate four representative data-agent harnesses:
$(i)$ \emph{DA-Agent}~\cite{huang2024code}, a data science agent equipped with Bash, Python, and SQL execution tools, reactively invoking tools with execution feedback and a moving memory window.
We set 80 maximum steps, and retrain the default 15-step history window and 1-minute step-level timeout;
$(ii)$ \emph{Smolagents}~\cite{smolagents}, a general-purpose ReAct-style agent that iteratively generates and executes code snippets with execution feedback and periodic planning.
We cap the number of coding steps at 40 to limit memory growth, and impose a 5-minute per-step timeout to handle unstable Jupyter Kernel Gateway connections;
$(iii)$ \emph{Claude Code}~\cite{claudecode} and \emph{CodeX}~\cite{codex}, two widely used ReAct-style agent harnesses supporting long-horizon planning, environment interaction (e.g., Bash, files, and coding), concurrent execution, and automatic context management.
We cap execution time at 60 minutes per task with an adaptive step-level timeout mechanism.
We pair each harness with each LLM in a compositional manner.

\hi{Diverse Domains.}
We include 15 real-world data science domains. Real-world business domains from Ant Group include: $(i)$ \textit{Financial}, involving cross-table aggregation of fund holdings, returns, and financial metrics; $(ii)$ \textit{Loan Model}, covering end-to-end credit risk modeling; $(iii)$ \textit{Loan Risk}, focusing on post-deployment model monitoring and metric-driven analysis; $(iv)$ \textit{Marketing}, targeting conversion rate prediction across businesses; and $(v)$ \textit{Strategy}, supporting business decision-making and multi-faceted strategy analysis.
Public domains include: $(vi)$ \textit{Agriculture}, involving correlated agricultural environments, production, and markets; $(vii)$ \textit{E-commerce}, covering products and user behaviors across major platforms; $(viii)$ \textit{Energy}, supporting cross-regional and temporal analysis of industrial consumption and energy indicators; $(ix)$ \textit{Entertainment}, capturing consumption of multimodal entertainment content; $(x)$ \textit{Healthcare}, comprising heterogeneous clinical, biomedical, and lifestyle data; $(xi)$ \textit{Real Estate}, integrating housing properties with socio-economic conditions; $(xii)$ \textit{Sports}, including records of teams, matches, and athletes across events; $(xiii)$ \textit{Social Network}, reflecting user behaviors and content across different platforms; $(xiv)$ \textit{Tourism}, enabling cross-country and travel-related analysis; and $(xv)$ \textit{Transportation}, modeling spatiotemporal mobility patterns in urban systems.

\hi{Datasets and Tasks.}
\oursys consists of a total of 344 tasks and 342 data files, amounting to 27.3 GB. As shown in \autoref{tab:dataset}, our carefully curated data pipeline enables \oursys to capture the complexity of real-world data science workflows, where each task involves, on average, 6.4 data files, 493.4 MB of data, and 23.5 skill applications. 

To illustrate skill-level characteristics across domains, we compute the ratio of skill categories within each domain, as shown in \autoref{fig:domain-skill}. We also identify the most representative and challenging skills per domain, summarized in \autoref{tab:domain-skill}.
Representative skills are determined using TF-IDF, calculated as skill frequency scaled by $\log(\text{domain tasks}/\text{skill tasks})$. Challenging skills are identified by assigning skill application scores via LLM for each task, aggregating them across domain tasks, and selecting the lowest-scoring skills.
We find that each domain exhibits distinct skill usage patterns, collectively covering a diverse range of data science task patterns. 
For example, the \textit{Marketing} domain emphasizes \textit{Data Modeling} skills, with prevalent use of ``Model Training and Customization'' to develop diverse models with rich feature representations for predicting key business indicators (e.g., user payment propensity).
\autoref{tab:domain-skill} further highlights the most challenging skills in this domain, where computing business metrics (e.g., conversion rates) over complex schema and modeling high-dimensional features frequently lead to failures of data agents.
In contrast, the \textit{Transportation} domain emphasizes \textit{Data Manipulation} skills, focusing on integrating heterogeneous data through transformations (e.g., ``Time Series Alignment and Matching'') to support spatiotemporal analysis.

\hi{Evaluation Metrics.}
We adopt a two-level evaluation. 
First, we implement five scoring functions to assess accuracy: table matching, modeling-based scoring, JSON matching, chart matching, and text matching, using Pass@1 by comparing data agent outputs with ground truth, tailored to data format and task type.
Second, we perform skill-level analysis by leveraging LLMs to identify misused skills for each task based on annotated skills, revealing failure patterns of data agents.
Technical details are provided in Section~\ref{subsec:pipeline}.

\subsection{Overall Performance Evaluation}
\label{subsec:overall}

We begin by evaluating each agent's performance on each dataset, with the overall results summarized in \autoref{tab:overall}.

\begin{figure*}[!t]\vspace{-1.5em}
  \centering
\includegraphics[width=\linewidth]{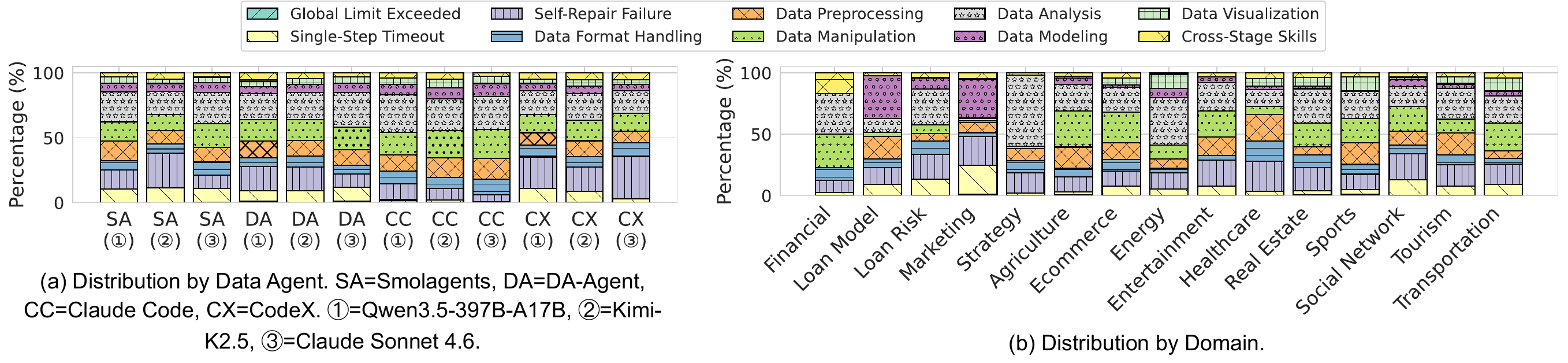}
  \vspace{-2.75em}
  \caption{Data Agent Failure Distribution.}
  \label{fig:error-overall}
  \vspace{-1.5em}
\end{figure*}

\hi{Agent Harness Comparison.}
We start by comparing different agent harnesses, and make two key observations.
First, among the evaluated data agents, the top three overall performers are \emph{CodeX (Kimi-K2.5)}, \emph{Smolagents (Qwen3.5)}, and \emph{Smolagents (Claude 4.6)}, suggesting that production-grade agent harnesses currently outperform the data science-specific \daagent.
This performance gap arises because \daagent adopts a lightweight design with limited engineering optimizations (e.g., a fixed memory horizon) and currently lacks specialized components, such as data profiling tools and data science–specific skills.
Second, we find that no agent harness achieves the best score across all domains, indicating that different harnesses have distinct domain-specific advantages.
For example, \smolagents performs best on the \textit{Marketing} domain, where many tasks involve large single data files ($\sim$1 GB or more).
\smolagents uses a notebook-based implementation that shares loaded data across steps for improved efficiency.
In contrast, other data agents often generate separate files at each step, repeatedly reloading the original data during exploration and execution, which can cause timeouts and sub-optimal performance (see also \autoref{fig:error-overall}b).
Besides, \daagent can also perform best on the \textit{Real Estate} domain.
This domain is challenging due to multi-source data alignment and complex metric calculations.
In this setting, \daagent generates large code blocks and achieves relatively high (though still low) scores, whereas other agents iteratively generate small code pieces, leading to inconsistent outputs.

\begin{table}[!t]
\centering
\caption{Average Metrics per Trajectory by Data Agent. SA=Smolagents, DA=DA-Agent, CC=Claude Code, CX=CodeX. \ding{192}=Qwen3.5-397B-A17B, \ding{193}=Kimi-K2.5, \ding{194}=Claude Sonnet 4.6.}
\vspace{-1em}
\label{tab:agent-metrics}
\resizebox{\linewidth}{!}{
\begin{tabular}{|c|c|c|c|c|c|}
\hline
\textbf{Data Agent} & \textbf{\# Steps} & \begin{tabular}[c]{@{}c@{}}\textbf{Tokens}\\\textbf{(K)}\end{tabular} & \begin{tabular}[c]{@{}c@{}}\textbf{Cost}\\\textbf{(\$)}\end{tabular} & \begin{tabular}[c]{@{}c@{}}\textbf{Success}\\\textbf{Steps (\%)}\end{tabular} & \begin{tabular}[c]{@{}c@{}}\textbf{Finish}\\\textbf{(\%)}\end{tabular} \\
\hline
SA (\ding{192}) & 18.1 & 319.4 & 0.07 & 94.6 & 100.0 \\
\hline
SA (\ding{193}) & 21.2 & 379.4 & 0.26 & 88.1 & 100.0 \\
\hline
SA (\ding{194}) & 24.0 & 493.5 & 1.66 & 97.0 & 99.7 \\
\hline
DA (\ding{192}) & 20.2 & 263.0 & 0.06 & 95.5 & 97.4 \\
\hline
DA (\ding{193}) & 17.1 & 145.4 & 0.11 & 94.1 & 98.8 \\
\hline
DA (\ding{194}) & 13.6 & 123.7 & 0.52 & 91.9 & 94.8 \\
\hline
CC (\ding{192}) & 28.1 & 683.4 & 0.12 & 93.1 & 99.7 \\
\hline
CC (\ding{193}) & 23.9 & 530.5 & 0.32 & 93.9 & 99.7 \\
\hline
CC (\ding{194}) & 16.8 & 408.3 & 0.47 & 96.7 & 99.7 \\
\hline
CX (\ding{192}) & 26.1 & 513.2 & 0.10 & 91.1 & 96.8 \\
\hline
CX (\ding{193}) & 40.2 & 1091.2 & 0.19 & 59.5 & 99.7 \\
\hline
CX (\ding{194}) & 20.0 & 218.7 & 0.76 & 92.2 & 88.1 \\
\hline
\end{tabular}
}
\vspace{-1.5em}
\end{table}

\hi{LLM Comparison.}
Next, we compare different LLMs within the same agent harness.
We find that the LLM achieving the best score varies across the four evaluated agent harnesses.
$(i)$ \claude performs best within \daagent and \claudecode.
This advantage is mainly due to the superior coding capabilities of \claude, which result in fewer syntax errors and more efficient, instruction-following code implementations.
For example, \claude is able to read Parquet files with specified columns, avoiding large-scale data loading and reducing the risk of timeouts.
$(ii)$ \qwen performs best within \smolagents.
This is because \kimi and \claude are less adaptable to the prompts of \smolagents, causing them to sometimes overlook parts of the instructions (e.g., wrapping code within \texttt{<code>} and \texttt{</code>}), which leads to repeated parsing errors.
$(iii)$ \kimi performs best within \codex.
This is because both \qwen and \claude are ill-suited to \codex: \qwen tends to generate responses misaligned with \codex (e.g., invalid function parameter errors, large blocks of inefficient code), while \claude frequently stops early without producing a task solution and fails to benefit from auto-compaction of memory (e.g., when the LLM context overflows).

\hi{Cost and Efficiency.}
We also record trajectory-level statistics, including token consumption and execution efficiency, as shown in \autoref{tab:agent-metrics}. We make two observations.
We have two observations.
First, although \emph{CodeX (Kimi-K2.5)} achieves the highest overall score, it exhibits the largest number of execution steps and the lowest successful-step ratio.
This is because \codex aggressively explores multiple solution paths and relies on rapid execution feedback to iteratively refine its trajectory.
Interestingly, although \emph{CodeX (Kimi-K2.5)} uses more tokens than \emph{Smolagents (Kimi-K2.5)}, its cost is lower. This is because \codex is more input-heavy and output-efficient, while output tokens are substantially more expensive than input tokens for \kimi.
Second, although \claude achieves the best performance within both \daagent and \claudecode, it is consistently more expensive than the open-source alternatives. We also observe that \claudecode appears particularly well optimized for Claude models, yielding higher token efficiency and cache utilization. For example, \emph{Claude Code (Claude 4.6)} achieves a higher score than \emph{Claude Code (Kimi-K2.5)} (46.6 vs 43.3) while costing only 1.5$\times$ more, compared to 4-6x in other harnesses.

\begin{figure}[!t]
  \centering
  \includegraphics[width=\linewidth]{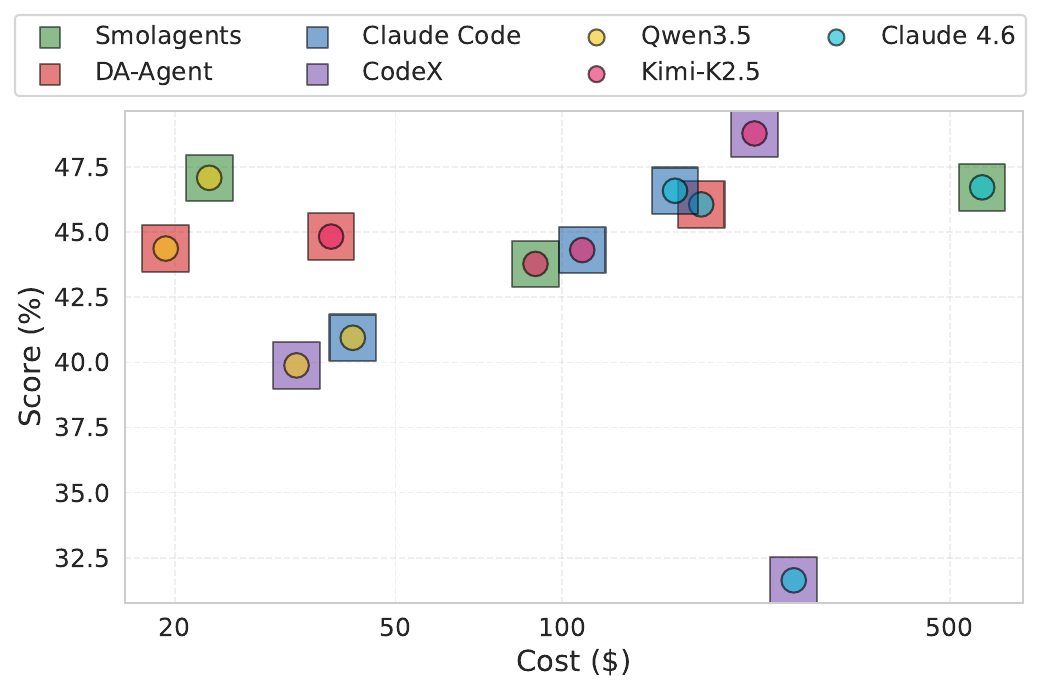}
  \vspace{-2.5em}
  \caption{Trade-off between Cost and Score.}
  \label{fig:tradeoff}  \vspace{-2em}
\end{figure}

We also visualize the trade-off between task score and token cost in \autoref{fig:tradeoff}. We make two observations.
First, \emph{Smolagents (Qwen3.5)} and \emph{DA-Agent (Qwen3.5)} achieve favorable cost-performance trade-offs, attaining moderate task scores at the lowest token costs. This is largely because \qwen is the least expensive evaluated LLM, while both \smolagents and \daagent adopt relatively lightweight orchestrations.
Second, \emph{Smolagents (Claude 4.6)} and \emph{CodeX (Claude 4.6)} exhibit comparatively poor cost-performance trade-offs, incurring substantially higher token costs without proportional score improvements. This observation is consistent with the LLM–harness mismatch discussed above.

\begin{figure*}[!t]\vspace{-1em}
  \centering
  \includegraphics[width=1.05\linewidth]{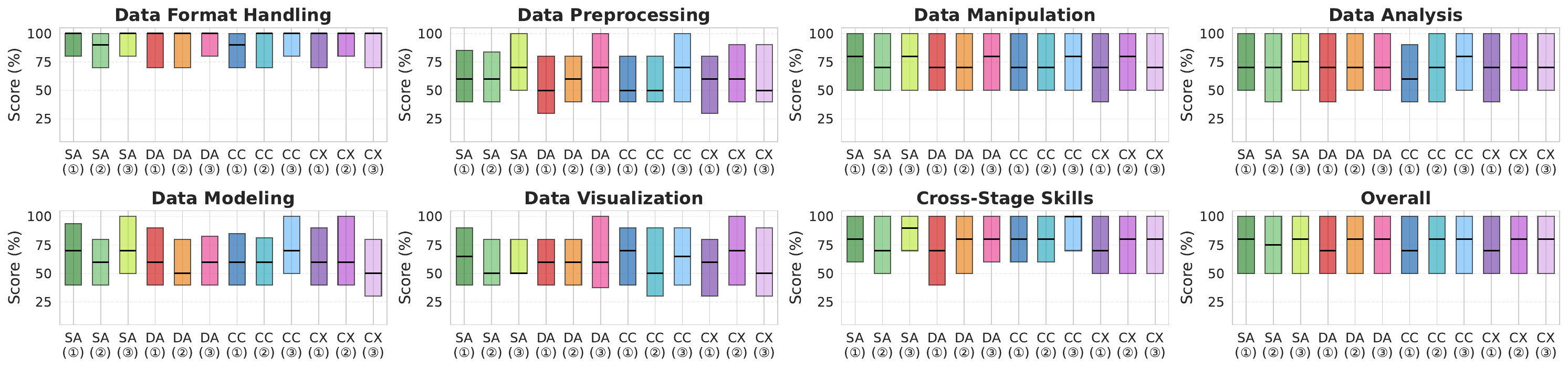}
  \vspace{-2em}
  \caption{Skill Score Comparison across Categories. SA=Smolagents, DA=DA-Agent, CC=Claude Code, CX=CodeX. \ding{192}=Qwen3.5-397B-A17B, \ding{193}=Kimi-K2.5, \ding{194}=Claude Sonnet 4.6.}
  \label{fig:skill-category}
  \vspace{-1.5em}
\end{figure*}

\begin{tcolorbox}[enhanced,sharp corners,
	width={8.5cm},
	colback=white,
	borderline={0.3mm}{0.3mm}{white},
	top=.5mm, bottom=.5mm, left=.5mm, right=.5mm]
Finding 1. {\it General-purpose harnesses achieve higher accuracy via mature components (high cost), e.g., CodeX $>$ Smolagents $>$ Claude Code $>$ DA-Agent. Smolagents also benefits from cross-step continuity. Higher accuracy generally comes at higher token costs, whereas lightweight harnesses (Smolagents and DA-Agent) can achieve favorable cost-performance trade-offs.}
\end{tcolorbox}

\begin{tcolorbox}[enhanced,sharp corners,
	width={8.5cm},
	colback=white,
	borderline={0.3mm}{0.3mm}{white},
	top=.5mm, bottom=.5mm, left=.5mm, right=.5mm] 
Finding 2. {\it In terms of token cost, Claude 4.6 incurs much higher token cost than open-source LLMs. In terms of accuracy, the best LLM varies across agent harnesses, emphasizing the adaptivity between LLM and harness (e.g., CodeX (Kimi-K2.5) benefits from more concurrent exploration steps).}
\end{tcolorbox}

\begin{figure*}[!t]
  \centering
  \includegraphics[width=\linewidth]{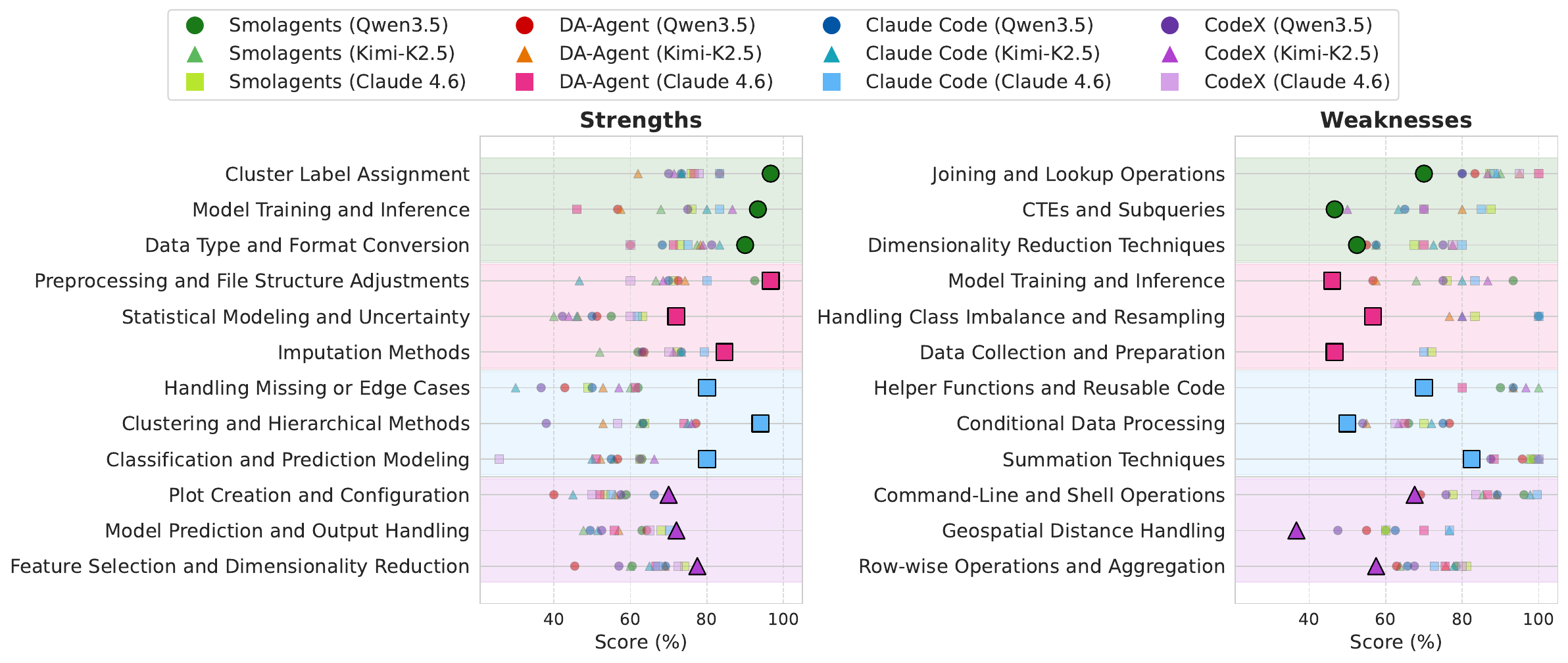}
  \vspace{-2.75em}
  \caption{Skill-Level Strengths and Weaknesses of the Top-Performing Agent per Harness: \emph{CodeX (Kimi-K2.5)}, \emph{Smolagents (Qwen3.5)}, \emph{Claude Code (Claude 4.6)}, and \emph{DA-Agent (Claude 4.6)}.}
  \label{fig:pro-con}
  \vspace{-1.75em}
\end{figure*}

\subsection{Skill-Level Performance Analysis}
\label{subsec:skill-exp}
Macro-level aggregate metrics do not reveal fine-grained root causes of data agent performance.
To enable deeper analysis of step-wise behaviors, we utilize skill annotations for each \oursys task. Since these skills represent common data-related operational patterns, they provide a breakdown of key steps in end-to-end tasks, enabling the evaluation of individual skill applications to uncover failure patterns in agent performance.
Specifically, by comparing data agent solutions with ground truth and leveraging feedback from evaluation functions, we use LLMs to score each skill application, where inefficient or incorrect executions receive lower scores aligned with the task evaluation score.
For each data agent and skill, we compute the skill score as the average score across all applications, and filter out skills with fewer than three applications to ensure reliability.

\hi{Skill Category Comparison.}
For each data agent, we aggregate skill scores to compute the average capability for each skill category, as shown in \autoref{fig:skill-category}.
We have two observations.
First, the overall skill scores are broadly consistent with task accuracy scores, supporting the validity of our skill-level evaluation.
For example, the relative ranking of LLMs is generally preserved within each agent harness.
Second, performance varies substantially across skill categories and data agents.
For example, \emph{Data Format Handling} and \emph{Cross-Stage Skills} generally achieve higher scores, because they are more closely aligned with common coding tasks encountered during pretraining and require less domain-specific data science knowledge.
Besides, \emph{CodeX (Qwen3.5)} performs particularly poorly on \emph{Cross-Stage Skills}, frequently producing erroneous shell commands (``Command-Line \& Shell Operations''; see also \autoref{fig:pro-con}) or failing to recover from missing dependencies (``Dependency Management''). These weaknesses partially explain its low performance on the \textit{Tourism} domain (see also \autoref{tab:domain-skill}).

\hi{Skills of Data Agents.}
To characterize fine-grained skill profiles,
we rank skills by their average score across all agents, weighted by $1-e^{-\text{skill frequency}/40}$ to emphasize frequent skills. The lowest-scoring skills are \textit{Data Alignment \& Merging}, \textit{Histogram Creation and Manipulation}, and \textit{Text Processing and Cleaning}, highlighting common limitations in processing heterogeneous and non-relational data.
For individual agents, we rank skills by the weighted deviation from the average score of all other agents.
Positive and negative values indicate strengths and weaknesses, respectively, as shown in \autoref{fig:pro-con}.
Due to space constraints, we report only the top-performing agent from each of the four harnesses, ranked by overall task accuracy.
We find that different data agents exhibit distinct strengths and weaknesses.
For example, all agents using \claude perform strongly on ``Statistical Modeling and Uncertainty'', suggesting that this capability is primarily determined by the underlying LLM.
\claude more faithfully follows task instructions and prefers installing and leveraging mature Python packages (e.g., using the t-distribution or ARIMA) over ad hoc implementations.
In contrast, all \daagent-based agents consistently underperform on ``Model Training and Inference'', because its static 1-minute per-step timeout can terminate training on high-dimensional feature tables, forcing agents to resort to simpler models.

\hi{Skills of Domains.}
To differentiate challenging scenarios across domains, we summarize the lowest-scoring skills for each domain in \autoref{tab:domain-skill}, and make two observations.
First, the lowest-scoring skills differ from the representative skills of the domain, focusing on challenging task execution (e.g., ``Time Series Analysis and Causality'') rather than data characteristics (e.g., ``Compression and Archiving'').
Second, both the skill scores and specific skills vary across domains, reflecting differences in complexity (e.g., the relatively easy \textit{Financial} domain with high scores) and task patterns.
\begin{tcolorbox}[enhanced,sharp corners,
	width={8.5cm},
	colback=white,
	borderline={0.3mm}{0.3mm}{white},
	top=.5mm, bottom=.5mm, left=.5mm, right=.5mm]
Finding 3. {\it Skill-level strength and weakness varies across agents, domains, and skill categories, partially aligning with task accuracy, with consistent weaknesses in processing heterogeneous and non-relational data overlooked by current harness designs.}
\end{tcolorbox}

\subsection{Failure Analysis of Data Agents}
\label{subsec:fail-analysis}
Our empirical analysis of data science tasks identifies ten primary categories of errors in data agent execution: $(i)$ \emph{Global Limit Exceeded}, where the agent exceeds global constraints on maximum steps or total runtime; $(ii)$ \emph{Single-Step Timeout}, where a single-step execution exceeds its time limit; $(iii)$ \emph{Self-Repair Failure}, where the agent fails to resolve errors from execution feedback; and $(iv)$ seven skill-specific error categories corresponding to the seven skill categories.
Our analysis results are shown in \autoref{fig:error-overall}.

\hi{Data Agent Comparison.}
We compare failure distributions across data agents in \autoref{fig:error-overall}a and make two observations.
First, \emph{Data Analysis} accounts for the largest proportion of failures, despite not being the most frequently invoked skill category.
These failures often arise from data validation, summarization, and statistical calculation, and may further propagate to downstream stages.
Second, high rates of \emph{Global Limit Exceeded}, \emph{Single-Step Timeout} and \emph{Self-Repair Failure} suggest poor adaptation between LLMs and agent harnesses, consistent with Section~\ref{subsec:overall}.
For example, \emph{CodeX (Qwen3.5)} exhibits a high rate of \emph{Self-Repair Failure}, frequently producing invalid function arguments or Bash commands with minor syntax errors (e.g., missing quotes or whitespace issues).

\hi{Domain Comparison.}
We compare failure modes across domains in \autoref{fig:error-overall}b and find that failure distributions vary substantially with data characteristics and task requirements. For example, the \emph{Marketing} domain exhibits the largest share of \emph{Global Limit Exceeded} and \emph{Single-Step Timeout} failures, primarily due to repeated loading of large-scale data files ($\sim$1 GB). The \emph{Healthcare} domain shows the highest rate of \emph{Self-Repair Failure}, as its heterogeneous data formats (e.g., ARFF) frequently trigger file parsing errors.
In contrast, the \emph{Loan Model} domain is most affected by \emph{Data Modeling} failures, owing to its simple file structure (two wide tables) and stronger reliance on complex feature derivation and modeling procedures, which are particularly challenging for current agents.

\hi{Ablation Study on Execution Budgets for ``Global Limit Exceeded'' and ``Single-Step Timeout'' Failures.}
We re-run the failed tasks due to global limit exceed and single-step timeout constraints, by increasing the global step/runtime limits and the per-step timeout (sampling up to 10 failed tasks), respectively.
We find that $(i)$ tasks exceeding the global limits account for less than 0.6\% of all tasks, and $(ii)$ neither intervention significantly improves task scores.
Instead, increasing the global limits merely prolongs unproductive execution loops, while increasing the per-step timeout can even mislead agents into less effective reasoning trajectories.

\begin{tcolorbox}[enhanced,sharp corners,
	width={8.5cm},
	colback=white,
	borderline={0.3mm}{0.3mm}{white},
	top=.5mm, bottom=.5mm, left=.5mm, right=.5mm]
Finding 4. {\it Data Analysis contributes the largest share of failures, whereas Global Limit Exceeded, Single-Step Timeout, and Self-Repair Failure reflect LLM–harness misalignment.
Failure modes also vary substantially across domains due to differences in data scale, file structure, and task complexity.}
\end{tcolorbox}

\section{Conclusion}

We propose a comprehensive benchmark for evaluating data agents, named \oursys, which covers realistic tasks spanning diverse domains with fine-grained labels.  We design a hierarchical skill extraction algorithm that leverages LLM-based semantic refinement to perform agglomerative clustering aligned with skill boundaries. We implement task selection and generation modules to ensure controlled skill coverage, enabling the inclusion of skill-diverse real-world tasks and realistic task simulations. Finally, through an in-depth empirical study of state-of-the-art data agents, we uncover key insights and identify important open problems to inspire future research. Our contributions pave the way for advancing the capabilities of autonomous data-science agents and fostering innovation in dynamic, skill-centered data-science systems.

\begin{acks}
We thank the following contributors for their support in providing and curating the business datasets used in this work: Yuyang Xia, Ziyu Jiang, Yingqi Gao, Xiongfeng Guo, Siyue Liu, Xinyu Li, Fengqin Wei, Xiaochen Liu, Chenlong Li, Haixia Peng, Minzhi Tang, Wenyi Liu, Mengzhen Zhang, Shan Zhang, Jieyuan Chen, Wenyan Liu, Xiuyun Yu, Fan Gou, Linyi Li, Siyu Lv, Shenkang Gu, and Linqi Li.
\end{acks}

\clearpage

\balance
\bibliographystyle{ACM-Reference-Format}
\bibliography{refs}
\balance

\end{document}